%% file: Sudakov.tex
\documentclass[12pt,bibliography=totoc,DIV=20]{scrartcl}
\pdfoutput=1

\input{paperdef}

\hypersetup{
  pdfauthor={Florian Domingo and Sebastian Paßehr}
  ,pdftitle={Electroweak corrections to the fermionic decays of heavy Higgs states}
  ,pdfsubject={}
  ,pdfkeywords={}
}

\begin{document}


\thispagestyle{empty}

\def\thefootnote{\fnsymbol{footnote}}

\begin{flushright}
BONN-TH-2019-02\\
\end{flushright}

\vfill

\begin{center}

\mytitle{
  Electroweak corrections \\to the fermionic decays \\of heavy Higgs states
}

\vspace{1cm}

Florian Domingo$^{1}$\footnote{email: florian.domingo@csic.es}
and
Sebastian Pa{\ss}ehr$^{2}$\footnote{email: passehr@lpthe.jussieu.fr}

\vspace*{1cm}

\textsl{
$^1$Bethe Center for Theoretical Physics \&
Physikalisches Institut der Universit\"at Bonn,\\
Nu\ss allee 12, D--53115 Bonn, Germany
}

\medskip
\textsl{
$^2$Sorbonne Université, CNRS,
Laboratoire de Physique Théorique et Hautes Énergies (LPTHE),\\
4 Place Jussieu, F--75252 Paris CEDEX~05, France
}

\end{center}

\vfill

\begin{abstract}{}
\input{00_Abstract}
\end{abstract}

\vfill

\def\thefootnote{\arabic{footnote}}
\setcounter{page}{0}
\setcounter{footnote}{0}
\newpage
\hypersetup{linkcolor=black}
\tableofcontents\label{TOC}
\hypersetup{linkcolor=blue}

\input{01_Introduction}
\input{02_THDM}
\input{03_NumericalAnalysis}
\input{04_Conclusions}

\section*{\tocref{Acknowledgments}}

We thank S.~Heinemeyer, M.~Stahlhofen and G.~Weiglein for useful
discussions. S.~P. acknowledges support by the~ANR grant
\mbox{``HiggsAutomator''} \mbox{(ANR-15-CE31-0002)}.

\begingroup
\let\secfnt\undefined
\newfont{\secfnt}{ptmb8t at 10pt}
\setstretch{.5}
\bibliographystyle{h-physrev}
\bibliography{literature}
\endgroup

\end{document}

%% file: paperdef.tex
\usepackage[T1]{fontenc}
\usepackage[utf8]{inputenc}

\usepackage{amsmath,amsfonts}
\usepackage[mathscr]{euscript}
\usepackage{booktabs}
\usepackage{cite}
\usepackage{graphicx, subfigure, ulem, hhline}
\usepackage[table]{xcolor}
\usepackage{siunitx}
\usepackage{soul}
\usepackage{wasysym}              
\usepackage{hyphenat}             

\usepackage{afterpage}
\usepackage{etoolbox,xstring,xspace,calc,xifthen}

\allowdisplaybreaks
\let\theparentequation\theequation
\patchcmd{\theparentequation}{equation}{parentequation}{}{}

\renewenvironment{subequations}[1][]{
  \refstepcounter{equation}%
  \setcounter{parentequation}{\value{equation}}
  \setcounter{equation}{0}
  \def\theequation{\theparentequation\alph{equation}}%
  \let\parentlabel\label
  \ifx\\#1\\\relax\else\label{#1}\fi
  \ignorespaces
}{%
  \setcounter{equation}{\value{parentequation}}
  \ignorespacesafterend
}

\newcommand*{\nextParentEquation}[1][]{
  \refstepcounter{parentequation}
  \setcounter{equation}{0}
  \ifx\\#1\\\relax\else\parentlabel{#1}\fi
}

\makeatletter
\renewcommand\@makefntext[1]{\leftskip=.8em\hskip-.62em\@makefnmark#1}
\makeatother

\usepackage{setspace}

\usepackage{jheppub}

\usepackage[all]{hypcap}
\usepackage{footnotebackref}

\usepackage[format=plain, margin=0.5cm, font=footnotesize, labelfont=bf]{caption}
\DeclareCaptionSubType*[arabic]{figure}
\DeclareCaptionSubType*[arabic]{table}
\let\theparentequation\theequation
\patchcmd{\theparentequation}{equation}{parentequation}{}{}

\apptocmd{\thebibliography}{\scriptsize}{}{}
\let\OLDthebibliography\thebibliography
\renewcommand\thebibliography[1]{
  \OLDthebibliography{#1}
  \setlength{\parskip}{1pt}
  \setlength{\itemsep}{1pt plus 0.3ex}
}

\newcommand{\citere}[1]{Ref.\,\cite{#1}}
\newcommand{\citeres}[1]{Refs.\,\cite{#1}}
\newcommand{\simord}{\mathord{\sim}\,}

\DeclareMathOperator{\gsim}{\raisebox{-.3em}{$\stackrel{\displaystyle >}{\sim}$}}

\newcommand{\Amp}[4][\mathcal{A}]{#1^{\mbox{\tiny #2}}_{\mbox{\tiny #3}}\ifthenelse{\isempty{#4}}{}{{\left[#4\right]}}}

\newcommand{\IE}{\textit{i.\,e.}}
\newcommand{\EG}{\textit{e.\,g.}}

\newcommand{\cp}{\ensuremath{{\cal CP}}}

\newcommand{\GamRat}[5]{\frac{\Gamma^{\text{#1}}_{#3}{\left[\mbox{#2}\right]}}{\Gamma^{\text{Born}}_{#5}{\left[\mbox{#4}\right]}}}
\newcommand{\dGamma}[3][]{\frac{\Delta\Gamma^{\text{#2}}_{#1}}{\Gamma^{\text{Born}}_{#1}}{\left[\mbox{#3}\right]}}
\newcommand{\AtoB}[2]{\mbox{$#1\to #2$}}

\newcommand{\DRbar}{\ensuremath{\overline{\mathrm{DR}}}}
\newcommand{\MSbar}{\ensuremath{\overline{\mathrm{MS}}}}

\makeatletter
\DeclareRobustCommand\em{%
  \@nomath\em \ifdim \fontdimen\@ne\font >\z@\scshape
  \else \slshape \fi}
\makeatother

\renewcommand{\emph}[1]{{\em #1}}

\addtokomafont{disposition}{\rmfamily}
\BeforeTOCHead[toc]{{\pdfbookmark[1]{\contentsname}{toc}}}

\usepackage{diagbox}

\usepackage{array}

\newcommand*{\mytitle}[1]{%
  \parbox{\linewidth}{\setstretch{1.5}\centering\Large\textsc{\textbf{\boldmath #1}}}
}



%% file: 00_Abstract.tex
Extensions of the Standard Model often come with additional, possibly
electroweakly charged Higgs states, the prototypal example being the
Two-Higgs-Doublet Model. While collider phenomenology
does not exclude the possibility for some of these new scalar fields
to be light, it is relatively natural to consider masses in the
multi-TeV~range, in which case the only remaining light Higgs boson automatically receives SM-like properties. The appearance of a hierarchy between the
new-physics states and the electroweak scale then leads to sizable
electroweak corrections, \EG\ in the decays of the heavy Higgs bosons,
which are dominated by effects of infrared type, namely
Sudakov~logarithms. Such radiative contributions obviously affect the
two-body decays, but should also be paired with the radiation of
electroweak gauge~bosons (or lighter Higgs bosons) for a consistent
picture at the one-loop order. Resummation of the
leading terms is also relatively easy to achieve. We re-visit these
questions in the specific case of the fermionic decays of heavy Higgs
particles in the Next-to-Minimal Supersymmetric Standard Model, in
particular pointing out the consequences of the three-body final
states for the branching ratios of the heavy scalars.

%% file: 01_Introduction.tex
\tocsection[\label{sec:intro}]{Introduction}

Many extensions of the Standard Model~(SM) imply the existence of an
extended Higgs sector. While the reality of a SM-like Higgs boson is
firmly established by the~experiments at the Large Hadron
Collider~(LHC)\,\cite{Aad:2012tfa,Chatrchyan:2012ufa,Aad:2015zhl}, the
status of other Higgs states remains largely speculative. At the
moment, the absence of conclusive signals for such new states has only
limited implications since, in many models, only a marginal portion of
the parameter space has been actually tested. In fact, light
electroweakly-charged scalar particles with a mass below that of the
top quark continue to be phenomenologically
viable\,\cite{Bahl:2018zmf}, even though such scenarios receive
constraints from multiple directions. The situation is much looser for
singlet-dominated states, such as predicted in the Next-to-Minimal
Supersymmetric Standard model~(NMSSM) for instance (see
\EG~\citere{Domingo:2015eea} for a discussion of the constraints from
Run\,I at the~LHC). On the other hand, it is tempting to retreat to
the (multi-)TeV~scale for the mass of the new Higgs bosons, because
then the SM-like characteristics of the observed state
(see \citeres{Khachatryan:2016vau,Sirunyan:2018koj,Aad:2019mbh}) are
almost automatically fulfilled, due to the decoupling properties of
the heavy Higgs particles (under the assumption of perturbative
couplings). In such a case, however, the presence of a comparatively
high scale introduces a hierarchy with respect to the electroweak
interactions that could lead to large radiative corrections, typically
in the form of Sudakov~logarithms---see
\EG~\citeres{Sudakov:1954sw,Kuhn:1999nn,Denner:2000jv} for a few
related references.

Below, we specialize in the decays of heavy Higgs bosons in the
particular case of the~NMSSM, although our analysis can be easily
extended to any other model including additional Higgs states, in
particular models based on a Two-Higgs-Doublet~Model~(THDM) of
type\,II. The~NMSSM\,\cite{Ellwanger:2009dp,Maniatis:2009re}---as any
supersymmetric~(SUSY)\,\cite{Nilles:1983ge,Haber:1984rc} extension of
the~SM---shields the mass of the~SM-like Higgs against large radiative
corrections from new physics at high-energy scales (\EG~the~GUT or
Planck~scale), suggesting a technically natural answer to
the~`Hierarchy~Problem'. Other motivations are the resolution of
the~`$\mu$~problem'\,\cite{Kim:1983dt} or the rich phenomenology of
the Higgs or neutralino sectors,
see \EG\ \cite{Ellwanger:2015uaz,Domingo:2015eea,Conte:2016zjp,Guchait:2016pes,Badziak:2016tzl,Das:2016eob,Cao:2016uwt,Das:2017tob,Muhlleitner:2017dkd,Ellwanger:2014hia,Chakraborty:2015xia,Kim:2015dpa,Carena:2018nlf,Titterton:2018pba,Cao:2016nix,Xiang:2016ndq,Cao:2018rix,Ellwanger:2018zxt,Domingo:2018ykx}
for a few recent discussions. The~NMSSM contains one pair of charged
and four additional neutral Higgs bosons beyond the SM-like state,
involving two new~\cp-even and two~\cp-odd degrees of freedom. Several
public tools propose an evaluation of the two-body Higgs decays. The
standard has long been a QCD-improved calculation: the decay routines
of \texttt{NMSSMTools}\,\cite{Ellwanger:2004xm,Ellwanger:2005dv,Domingo:2015qaa,NMSSMTOOLS-www}
and \texttt{NMSSMCALC}\,\cite{Baglio:2013iia,NMSSMCALC-www} are based
on \texttt{HDECAY}\,\cite{Djouadi:1997yw,Djouadi:2018xqq};
\texttt{SOFTSUSY}\,\cite{Allanach:2001kg,Allanach:2013kza,Allanach:2017hcf}
also performs at the same order. More recently, full one-loop analyses
of the two-body Higgs decays have been performed with the
code~\texttt{SloopS}\,\cite{sloops-www,Belanger:2014roa,Belanger:2016tqb,Belanger:2017rgu}
or in the~\DRbar~$\left(\MSbar\right)$~scheme for generic
models\,\cite{Goodsell:2017pdq} with the
code \texttt{SPHENO}\,\cite{Porod:2003um,Porod:2011nf,Goodsell:2014bna,spheno-www},
which
employs~\texttt{SARAH}\,\cite{Staub:2009bi,Staub:2010jh,Staub:2012pb,Staub:2013tta}. Also
in \citere{Domingo:2018uim} the two-body Higgs decays into SM~final
states were considered at the full one-loop order. Similar projects
have been presented for the~THDM and its
extensions\,\cite{Krause:2018wmo,Krause:2019oar,Kanemura:2019kjg}. Yet---with
the exception of photon and gluon radiation as well as the production
of off-shell gauge~bosons below threshold that subsequently decay into
fermions (see \EG\ \citere{Spira:2016ztx} for a recent summary in
the~MSSM)---little attention has been paid to the three-body decays,
which, however, intervene at the same order as the one-loop
corrections to two-body decays.

The degeneracy among the states forming
an~$SU(2)_{\text{L}}$~multiplet is lifted by the electroweak symmetry
breaking, so that the mass-squared differences among the partners of a
scalar doublet are expected to be of the order of~$M_Z^2$. Therefore,
in scenarios of a~THDM where one doublet mass is much larger than the
electroweak scale, actually four Higgs bosons have almost degenerate
large masses and approximately organize as
an~$SU(2)_{\text{L}}$~doublet, while one Higgs boson is automatically
endowed with SM-like properties, thus making it a good candidate for
explaining the signals observed by the~ATLAS
and~CMS~collaborations---provided its mass falls within the suitable
interval at~$\simord125$\,GeV. This setup fulfills the conditions of
the decoupling limit\,\cite{Gunion:2002zf} and the decays of the heavy
Higgs bosons into~SM~final~states tend to be dominated by the
fermionic channels: the couplings of the heavy states to a pair of
electroweak gauge bosons vanish (approximately for the~\cp-even
Higgs), because the doublet formed by the heavy states is orthogonal
to that generating the electroweak-symmetry-breaking vacuum
expectation value~(v.e.v.); in addition, the decays into a~SM-like
Higgs plus an electroweak gauge~boson are also suppressed, because the
decoupling states are approximately partners of one another under the
electroweak gauge group, and not of the~SM-like Higgs;\footnote{Decays
of a heavy doublet into another heavy doublet state plus a gauge boson
are kinematically inaccessible, in general. However, production of an
off-shell gauge~boson decaying into fermions has been considered \EG\
in\,\cite{Djouadi:1995gv}.} finally, Higgs-to-Higgs decays are
suppressed in the limit of heavy initial states---the triple Higgs
coupling of electroweak size implies a suppression~$\simord M_Z/M_h$,
where~$M_h$ represents the mass of the initial Higgs boson. On the
other hand, at least some of the fermionic decays are expected to be
unsuppressed, depending on the type of~THDM. For a type\,II framework,
the standard search channels at the~LHC thus involve tau-onic
final~states, while the~$b\bar{b}$ or~$t\bar{t}$~decays are actually
sizable but suffer from the~QCD~background. The decays of
singlet-dominated states are more difficult to characterize
generically. A pure singlet could only decay radiatively into
SM~final~states (and even not at all for~\mbox{$\lambda=0$}). In
general, the decays of mostly singlet-like states into SM~particles
are thus dominated by their subleading doublet components---therefore
their bosonic decays, \EG\ into light Higgs bosons, are not
necessarily suppressed. Below, we focus on the fermionic decays for
simplicity, but the procedure can be generalized to bosonic
final~states as well.

It was observed in \citere{Domingo:2018uim} that the decays of heavy
new-physics Higgs states into SM~particles could receive sizable
radiative corrections beyond the well-known
QCD~effects\,\cite{Braaten:1980yq,Drees:1990dq}. It is desirable to
control such corrections for a better characterization of the expected
signals at colliders and a more quantitative implementation of
associated limits. At the~LHC, the expected reach does not exceed
masses of~$\simord 2$\,TeV\,\cite{CMS:2016DP}. However, already for
masses in the~TeV~range, so-called electroweak Sudakov
double~logarithms
\begin{gather}
\simord\frac{g_2^2}{16\pi^2}\ln^2\frac{M_V^2}{M_h^2}
\end{gather}
(where~$g_2$ represents the gauge coupling, and~$M_V$
and~\mbox{$M_h\sim1$--$2$}\,TeV denote the masses of the gauge and
heavy Higgs bosons respectively) attain the level of~$\simord10\%$.

The main purpose of this paper consists in analyzing the electroweak
corrections to the two-body fermionic decays of heavy Higgs bosons,
and their interplay with three-body decays involving the radiation of
an electroweak gauge boson. The noteworthy difference compared to the
case of~QED and QCD~corrections is that the radiation of massive
electroweak gauge bosons leads to clearly distinguishable final
states, as opposed to soft and collinear photons and gluons. The
corresponding decays could thus be measured separately and there is no
justification on the theoretical side for considering only inclusive
decay widths (summing over two-body decays and the corresponding ones
with radiated~$W,Z$): therefore the potentially large Sudakov
double~logarithms are experimentally accessible effects that are
expected to reduce the branching fractions of the two-body decays. In
any case, from order-counting it appears mandatory to take the
three-body final states into account if one wants to perform a
consistent analysis of the branching ratios for two-body decays of
heavy Higgs bosons at the full one-loop order. To our knowledge, such
effects have not been considered in the~NMSSM (or the~MSSM) yet, and
we expose here how we implement these channels in view of a future
inclusion within a version
of~\texttt{FeynHiggs}\,\cite{Heinemeyer:1998np,Heinemeyer:1998yj,Degrassi:2002fi,Frank:2006yh,Hahn:2013ria,Bahl:2016brp,Bahl:2017aev,FH-www}
dedicated to
the~NMSSM\,\cite{Drechsel:2016jdg,Domingo:2017rhb,Domingo:2018uim}.

In the following section, we discuss the formal aspects of our
evaluation of the two- and three-body decays, emphasizing the
possibility to capture most of the electroweak corrections within a
simple resummation of
Sudakov~double~logarithms\,\cite{Fadin:1999bq}. We then illustrate the
numerical impact of the electroweak corrections in a scenario with
heavy SUSY sector and focusing on doublet-dominated Higgs bosons in
the initial state. Finally, Sect.\,\ref{sec:conclusion} summarizes our
achievements.

%% file: 02_THDM.tex
\tocsection[\label{sec:theory}]{Higgs decays into SM fermions in the NMSSM}

In this section, we summarize the main theoretical ingredients
entering the decays of heavy Higgs bosons into SM~fermions. We follow
the notations of \citeres{Domingo:2017rhb,Domingo:2018uim}.

\tocsubsection{Two-body decay width}

\subsubsection*{Decay of the Higgs bosons at one-loop order: automated calculation}

We already described our full one-loop implementation of the two-body
fermionic decays of neutral Higgs bosons
in \citere{Domingo:2018uim}. For the sake of completeness, we
summarize the main ingredients in the following:
\begin{itemize}
\item
  We rely on the automated calculation
  of~\texttt{FeynArts}\,\cite{Kublbeck:1990xc,Hahn:2000kx}, \texttt{FormCalc}\,\cite{Hahn:1998yk}
  and~\texttt{LoopTools}\,\cite{Hahn:1998yk}, using the model~file
  (and renormalization scheme) presented in \citere{Domingo:2017rhb}.
\item
  External Higgs fields are upgraded to on-shell fields using the
  transformation matrix~$Z^{\mbox{\tiny mix}}$ that is determined by
  the LSZ~reduction, see \citere{Domingo:2017rhb}. The mixing of the
  external Higgs leg with the electroweak neutral current is processed
  diagrammatically, and we subtract gauge-violating effects of
  two-loop order that appear due to the difference between the
  (kinematical) loop-corrected Higgs mass and the tree-level mass,
  see \citere{Domingo:2018uim}. The fermion fields are renormalized
  on-shell.
\item
  The virtual QCD~and~QED~corrections are processed separately and
  combined with the real-radiation contributions in order to define an
  infrared~(IR)-finite correction
  factor\,\cite{Braaten:1980yq,Drees:1990dq,Dabelstein:1995js} to the
  pure born width~$\Gamma^{\text{Born}}$ describing the inclusive
  fermionic width (with radiated gluons and photons in the final
  state). The QCD~logarithms are absorbed within the definition of
  running Yukawa couplings evaluated at the scale of the decaying
  Higgs:
  \begin{align}
    \frac{\Gamma^{\text{QCD+QED}}}{\Gamma^{\text{Born}}} &=
    \left(\frac{Y_f{(M_h)}}{Y_f{(m_t)}}\right)^2
    \left[1+c^{\text{QCD}}+c^{\text{QED}}\right]
  \end{align}
  where~$Y_f$ represents the relevant Yukawa coupling absorbing
  QCD~logarithms; $c^{\text{QCD}}$ and~$c^{\text{QED}}$ represent
  the~QCD and QED~correction
  factors---see \EG~Eq.\,(\href{https://arxiv.org/pdf/1807.06322.pdf#equation.2.12}{2.12})
  of \citere{Domingo:2018uim}.
\item
  The leading SUSY~corrections amount to contributions to effective
  dimension~$4$ Higgs--fermion operators. They are explicitly
  extracted in the~$SU(2)$-conserving limit and re-arranged so as to
  provide a resummation of the~$\tan\beta$-enhanced effects,
  see \citeres{Banks:1987iu,Hall:1993gn,Hempfling:1993kv,Carena:1994bv,Carena:1999py,Eberl:1999he,Degrassi:2000qf,Buras:2002vd,Guasch:2003cv,Noth:2008tw,Noth:2010jy,Williams:2011bu,Baglio:2013iia,Ghezzi:2017enb};
  a linearized (non-resummed) version is subtracted from the actual
  one-loop diagrammatic calculation in order to avoid double~counting.
\item
  The full one-loop corrections (excluding the~QED and
  QCD~corrections) to the Higgs--fermion vertex are derived
  automatically from our model~file. However, we subtract
  gauge-violating effects of two-loop order through a re-definition of
  the Higgs--Goldstone~couplings by a shift of one-loop order, as
  explained in
  section \href{https://arxiv.org/pdf/1807.06322.pdf#section.2.1}{2.1}
  of \citere{Domingo:2018uim}.
\end{itemize}

Obviously, the two-body decays of the charged Higgs can be implemented
in a similar fashion. In particular:
\begin{itemize}
\item
  The external charged Higgs field and its mass are renormalized
  on-shell, thus requiring no normalization factor. The mixing with
  the electroweak charged current is processed diagrammatically.
\item
  The~QCD and QED~corrections are factorized out and combined with the
  real radiation process. The QCD-correction~factor is
  well-known\,\cite{Djouadi:1994gf}. The QED~contribution is somewhat
  more involved, due to the initial state being electrically
  charged. We compute this correction factor by considering separately
  the virtual piece as well as the soft and hard radiation. For the
  hard radiation, we explicitly take the limit~\mbox{$m_b\to0$}. For
  simplicity, we only provide the leading logarithmic terms
  for~\AtoB{H^-}{b\bar{t}} below:
  \begin{subequations}
  \begin{align}
    \frac{\Delta\Gamma^{\text{QED}}}{\Gamma^{\text{Born}}}[\AtoB{H^-}{b\bar{t}\,}]&=
    \frac{e^2}{16\,\pi^2}\left[C^{\text{virt}}+C^{\text{soft}}+C^{\text{hard}}\right],\\
    C^{\text{virt}}&=2 \left[
      Q_b^2\,\ln\frac{m_b^2}{M_{H^{\pm}}^2} +
      Q_t^2\,\ln\frac{m_t^2}{M_{H^{\pm}}^2} +
      2\left(1+Q_b\,Q_t\right)\right]
      \ln\frac{M_{H^{\pm}}^2}{m_{\gamma}^2}\,,\\
    \begin{split}
      C^{\text{soft}}&=2 \left[
        Q_b^2\,\ln\frac{m_b^2}{M_{H^{\pm}}^2} +
        Q_t^2\,\ln\frac{m_t^2}{M_{H^{\pm}}^2} +
        2\left(1+Q_b\,Q_t\right)\right]
        \ln\frac{m_{\gamma}^2}{\Omega^2_{\text{IR}}}\\
      &\quad-Q_b^2 \left[
        \ln^2\frac{m_b^2}{M_{H^{\pm}}^2} + 2\,\ln\frac{m_b^2}{M_{H^{\pm}}^2}\right]
      -Q_t^2 \left[
        \ln^2\frac{m_t^2}{M_{H^{\pm}}^2} + 2\,\ln\frac{m_t^2}{M_{H^{\pm}}^2}\right],
    \end{split}\\
    \begin{split}
      C^{\text{hard}}&=2 \left[
        Q_b^2\,\ln\frac{m_b^2}{M_{H^{\pm}}^2} +
        Q_t^2\,\ln\frac{m_t^2}{M_{H^{\pm}}^2} +
        2\left(1+Q_b\,Q_t\right)\right]
        \ln\frac{\Omega^2_{\text{IR}}}{M_{H^{\pm}}^2}\\
      &\quad+3\left[
        Q_b^2\,\ln\frac{m_b^2}{M_{H^{\pm}}^2} +
        Q_t^2\,\ln\frac{m_t^2}{M_{H^{\pm}}^2}\right],
    \end{split}
  \end{align}
  \end{subequations}
  where~$e$~represents the electric charge, $Q_{t,b}$~the fermion
  charges, $m_{\gamma}$~the photon-mass regulator,
  and~$\Omega_{\text{IR}}$ the~IR~cut on the photon energy.
\item
  Non-QCD and non-QED~one-loop diagrams are calculated with our
  automated procedure.
\end{itemize}

Below, we denote the decay widths that are derived in this automated
fashion at fixed order as~$\Gamma^{\text{FO}}$, while the Born results
read~$\Gamma^{\text{Born}}$. As in \citere{Domingo:2018uim}, the
renormalization scale is set equal to the pole top
mass~\mbox{$m_t=173.2$\,GeV} for all \DRbar-renormalized
counterterms. As we wish to focus on the electroweak corrections, we
will in practice consider a scenario with decoupling (heavy)
SUSY~particles. In this context, the SUSY~corrections essentially
reduce to their contributions to the dimension~$4$ effective
operators---see \citeres{Banks:1987iu,Hall:1993gn,Hempfling:1993kv,Carena:1994bv,Carena:1999py,Eberl:1999he,Degrassi:2000qf,Buras:2002vd,Guasch:2003cv,Noth:2008tw,Noth:2010jy,Williams:2011bu,Baglio:2013iia,Ghezzi:2017enb}---and
we thus fully factorize them as corresponding to a short-distance
effect.\footnote{Equivalently, it would be possible to factorize only
the contributions to dimension~$4$ operators. For commodity in the
processing of diagrams, it is more convenient to us to factor out the
full SUSY contribution. The difference between the two procedures is a
formally and numerically subleading term in the considered scenario.}
In this procedure, however, also the SUSY~contributions to the Higgs
wave~function---\IE\ intervening in the normalization of
the~$Z^{\text{mix}}$ matrix---need to be extracted accordingly. This
effective tree-level width, including SUSY~corrections, is later
denoted as~$\Gamma^{\text{eff}}$. Then, in \citere{Domingo:2018uim},
we kept the term of the squared one-loop amplitude in the definition
of the width: this was justified in the case where one-loop
contributions dominate over the tree-level result. However, the
electroweak corrections that we study below are essentially
proportional to the tree-level amplitude, meaning that they would be
suppressed together with it. In this case, the full one-loop squared
term is actually misleading, since its impact can be substantially
altered by the inclusion of the interference of the amplitude of
two-loop order with the tree-level one. We thus discard this term,
keeping only a squared one-loop amplitude for those contributions that
are not proportional to the tree-level term (and subleading in the
channels that we are considering).

\subsubsection*{Electroweak double logarithms}

In this one-loop evaluation of the two-body fermionic decays of heavy
Higgs bosons, several pieces of the radiative corrections beyond
the~QCD and~QED~contributions can matter, in general. However, in the
limit of heavy SUSY~spectra, most of the SUSY-related corrections are
logarithmic and can be captured within effective Yukawa couplings, as
mentioned above. As can be observed \EG\ in
Fig.\,\href{https://arxiv.org/pdf/1807.06322.pdf#figure.caption.245}{1}
of
\citere{Domingo:2018uim} (by comparing the green and red curves), this
effective description, which works reasonably well for a light Higgs,
still leaves out sizable one-loop effects for a heavy state. These
remaining one-loop contributions (reaching~$\simord10\%$ for
the~$1$\,TeV~states of
Fig.\,\href{https://arxiv.org/pdf/1807.06322.pdf#figure.caption.245}{1}
of \citere{Domingo:2018uim}) are associated with electroweak
corrections. Below, we explain how most of these relatively large
effects can be easily put under control.

\begin{figure}[b!]
  \centering
  \includegraphics[width=.7\textwidth]{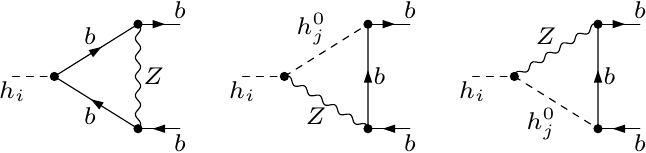}\\[2ex]
  \includegraphics[width=.7\textwidth]{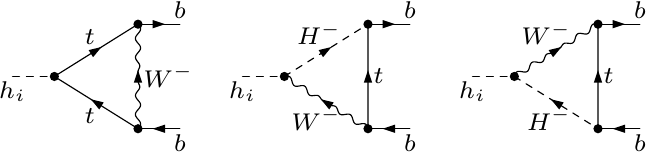}
  \caption{\label{fig:sud_diagrams} Contributions
  to~\AtoB{h_i}{b\bar{b}} leading to double logarithms.}
\end{figure}

The sizable electroweak contribution originates in the hierarchy
between the heavy Higgs state and the light particles entering in the
loops, which results in large logarithms of Sudakov~type. In
particular, for gauge interactions we expect to find double
logarithms~$\simord
g_2^2/\left(16\,\pi^2\right)\ln^2{\left(M_Z^2/M_h^2\right)}$.
The diagrams leading to double logarithms in the example
of~\AtoB{h_i}{b\bar{b}} are displayed in Fig.\,\ref{fig:sud_diagrams}
and include two topologies:
\begin{enumerate}
\item
  the well-known fermion--fermion--vector triangle:\\
  the associated loop-function
  contributes~$\simord\frac{1}{2}\ln^2{\left(M_V^2/M_h^2\right)}$ at
  the double-logarithmic order;
\item
  the fermion--scalar--vector triangle:\\  
  under the condition that the internal scalar is almost degenerate
  with the external Higgs, the loop function
  contributes~$\simord{-}\frac{1}{4}\ln^2{\left(M_V^2/M_h^2\right)}$;
  here, `almost degenerate' covers a range larger than~\mbox{$M_h\pm
  M_V$}, but much narrower than~$\left[M_h/2,\,2\,M_h\right]$.
\end{enumerate}
In our case, one cannot rely on the popular evaluations of the
double-logarithmic coefficient presented in
\citeres{Denner:2000jv,Fadin:1999bq}, amounting to a sum of the
quantity
\begin{gather}\label{eq:pop2L}
  -\frac{g_2^2}{32\,\pi^2}
  \left[I\left(I+1\right) + \left(Y/2\right)^2\,t_{\text{w}}^2
  - Q^2\,s_{\text{w}}^2\right]
  \ln^2{\frac{M_V^2}{M_h^2}}
\end{gather}
over the external legs---where~$I$, $Y$, $Q$ represent the isospin,
hypercharge and electric charge of the external state, $t_{\text{w}}$
and~$s_{\text{w}}$ denote the tangent and sine of the electroweak
mixing angle; the~$Q^2$~term subtracts the QED~contribution that is
considered separately. The reason why this formula fails is simply
that one of the (electroweakly charged) external states, the Higgs
line, is massive, actually determining the center-of-mass~energy.

For a doublet state~$h_i$ or~$H^{\pm}$ in the decoupling limit, the
electroweak partners of the external Higgs bosons are indeed always
almost degenerate with the external state and it is easy to extract
the Sudakov double~logarithms explicitly (with~$g_{1,2}$ the
electroweak couplings; for~\AtoB{H^-}{t\bar{b}}, the subscripts~$t,b$
refer to the pieces of the width that are proportional to the squared
Yukawa couplings~$Y_{t,\,b}^2$):
\begin{subequations}\label{eq:2bDL}
\begin{align}
  \dGamma{DL}{\AtoB{h_i}{b\bar{b}\,}} &\simeq -\frac{1}{48\,\pi^2}\left[
    \frac{5}{12}\,g_1^2 + \frac{9}{4}\,g_2^2 - \frac{2}{3}\,e^2\right]
  \ln^2\frac{M_V^2}{M_{h_i}^2}\,,\\
  \dGamma{DL}{\AtoB{h_i}{t\bar{t}\,}} &\simeq -\frac{1}{48\,\pi^2}\left[
    \frac{17}{12}\,g_1^2 + \frac{9}{4}\,g_2^2 - \frac{8}{3}\,e^2\right]
  \ln^2\frac{M_V^2}{M_{h_i}^2}\,,\\
  \dGamma{DL}{\AtoB{h_i}{\tau^+\tau^-}} &\simeq -\frac{1}{48\,\pi^2}\left[
    \frac{15}{4}\,g_1^2 + \frac{9}{4}\,g_2^2 - 6\,e^2\right]
  \ln^2\frac{M_V^2}{M_{h_i}^2}\,,\\
  \dGamma[b]{DL}{\AtoB{H^-}{t\bar{b}\,}} &\simeq -\frac{1}{48\,\pi^2}\left[
    \frac{5}{12}\,g_1^2 + \frac{9}{4}\,g_2^2 - \frac{5}{3}\,e^2\right]
  \ln^2\frac{M_V^2}{M_{H^{\pm}}^2}\,,\\
  \dGamma[t]{DL}{\AtoB{H^-}{t\bar{b}\,}} &\simeq -\frac{1}{48\,\pi^2}\left[
    \frac{17}{12}\,g_1^2 + \frac{9}{4}\,g_2^2 - \frac{5}{3}\,e^2\right]
  \ln^2\frac{M_V^2}{M_{H^{\pm}}^2}\,,\\
  \dGamma{DL}{\AtoB{H^-}{\tau^-\bar{\nu}_{\tau}}} &\simeq -\frac{1}{48\,\pi^2}
  \left[
    \frac{15}{4}\,g_1^2 + \frac{9}{4}\,g_2^2 - 3\,e^2\right]
  \ln^2\frac{M_V^2}{M_{H^{\pm}}^2}\,.
\end{align}
\end{subequations}
Here again, the terms~$\mathord{\propto}\,e^2$ simply correspond to
the QED~corrections, which are subtracted since they are
separately processed as a long-distance effect. We stress that the
double-logarithmic corrections always interfere destructively with the
tree-level amplitude, thus leading to a systematic reduction of
the two-body decay width.

For a singlet-dominated state, the situation is somewhat more
subtle. If the mass of this Higgs boson is far from that of other
Higgs states, then the fermion--scalar--vector triangle does not
produce double-logarithms for lack of a degenerate electroweak partner
of the external state. In this case, the double-logarithmic terms
coincide with the formula of Eq.\,\eqref{eq:pop2L}, applied to the
fermionic final states (one being left-handed, the other
right-handed). However, if the singlet finds itself accidentally in
the window of mass degeneracy with the doublet states, then the
fermion--scalar--vector topology is relevant, together with possibly
substantial doublet-singlet mixing.

For Higgs masses in the range of a~few~TeV, we expect electroweak
corrections to remain at the level of~$\simord 10\%$: this is still
perturbative and there is no deep call for resumming these
corrections. This only becomes necessary if one desires to extend the
result to masses as high as~$\mathcal{O}{\left(100\right)}$\,TeV. On
the other hand, it is remarkable that the leading double-logarithmic
terms can be controlled at all orders. Such a resummation also allows
us to define an effective tree-level result capturing the bulk of the
radiative effects. The resummation of double~logarithms can be
performed by considering the IR~behavior of the matrix element,
see \citere{Fadin:1999bq}: it eventually amounts to exponentiating the
double-logarithmic coefficient obtained at the one-loop order. We
stress that there is no complication from the separate treatment of
the QED~effects at the considered order. We can define the
`electroweakly~improved' Born~width~$\Gamma^{\text{EW}}$ as
\begin{align}
  \frac{\Gamma^{\text{EW}}}{\Gamma^{\text{Born}}} &=
  \exp\left[\frac{\Delta\Gamma^{\text{DL}}}{\Gamma^{\text{Born}}}\right].
\end{align}
In the \cp-violating case, there is no difficulty in processing the
scalar and pseudoscalar components separately, since the corresponding
operators do not interfere.

\subsubsection*{Single logarithms}

While we expect the double~logarithms to represent the numerically
largest piece of the electroweak corrections, it is relatively easy to
extract single logarithms as well. The latter originate in the vertex
corrections of Fig.\,\ref{fig:sud_diagrams}, but also in the
counterterm and wave-function normalization~$Z^{\mbox{\tiny mix}}$.
Obviously, they depend on the chosen scheme, \IE~the definition of the
tree-level Yukawa couplings. In our `on-shell' definition and with the
renormalization scale of the loop functions set to~\mbox{$m_t\sim
M_Z$} (following the prescriptions of \texttt{FeynHiggs}), we obtain
the following terms for the decays of doublet states
(neglecting~$Y_{\tau}^2$):
\begin{subequations}\label{eq:2bSL}
\begin{align}
  \dGamma{SL}{\AtoB{h_i}{b\bar{b}\,}} &\simeq \frac{1}{16\,\pi^2}\left[
    3\, Y_b^2 \left(1 + c_{\beta}^2\right) - Y_t^2 \left(2 + c_{\beta}^2\right)\right] \ln\frac{M_{h_i}^2}{M_Z^2}\,,\\
  \dGamma{SL}{\AtoB{h_i}{t\bar{t}\,}} &\simeq \frac{1}{16\,\pi^2}\left[
    3\, Y_t^2 \left(1 + s_{\beta}^2\right) - Y_b^2 \left(2 + s_{\beta}^2\right)\right] \ln\frac{M_{h_i}^2}{M_Z^2}\,,\\
  \dGamma{SL}{\AtoB{h_i}{\tau^+\tau^-}} &\simeq \frac{1}{16\,\pi^2}\left[
    3\, Y_b^2 \left(1 + c_{\beta}^2\right) - 3\, Y_t^2\, c_{\beta}^2\right] \ln\frac{M_{h_i}^2}{M_Z^2}\,,\\
  \dGamma[b]{SL}{\AtoB{H^-}{t\bar{b}\,}} &\simeq \frac{1}{16\,\pi^2}\left[
    3\, Y_b^2 \left(1 + c_{\beta}^2\right) - Y_t^2 \left(2 + c_{\beta}^2\right)\right] \ln\frac{M_{H^{\pm}}^2}{M_Z^2}\,,\\
  \dGamma[t]{SL}{\AtoB{H^-}{t\bar{b}\,}} &\simeq \frac{1}{16\,\pi^2}\left[
    3\, Y_t^2 \left(1 + s_{\beta}^2\right) - Y_b^2 \left(2 + s_{\beta}^2\right)\right] \ln\frac{M_{H^{\pm}}^2}{M_Z^2}\,,\\
  \dGamma{SL}{\AtoB{H^-}{\tau^-\bar{\nu}_{\tau}}} &\simeq \frac{1}{16\,\pi^2}\left[
  3\, Y_b^2 \left(1 + c_{\beta}^2\right) - 3\, Y_t^2\, c_{\beta}^2\right]\ln\frac{M_{H^{\pm}}^2}{M_Z^2}\,.
\end{align}
\end{subequations}
We note that the terms proportional to gauge couplings cancel out
(between the vertex contributions and the Higgs wave-function
correction), leaving only terms proportional to the Yukawa
couplings. The single logarithms are both of ultra-violet~(UV) and
IR~origins. A resummation would disentangle both types, attributing
the UV~logarithms to the running of the parameters. In practice,
however, we observe
that~\mbox{$\ln{\left(M_h^2/M_Z^2\right)}\lesssim\pi^2$}, so that
constant pieces (\IE~not depending on~$M_h$, as opposed to the double-
and single-logarithmic terms, in an expansion at $M_h\to\infty$)
compete with the single logarithms, making a resummation of the latter
superfluous.

\subsubsection*{`Improved' width}

We schematically summarize the expression of the `improved' prediction
of the two-body decay widths with resummed Sudakov double~logarithms
and factorized SUSY~corrections by the following equation:
\begin{align}\label{eq:impwidth}
  \Gamma^{\text{imp}}
  &= \frac{\Gamma^{\text{QCD+QED}}}{\Gamma^{\text{Born}}} \left\{ \Gamma^{\text{FO}}
  + \Gamma^{\text{eff}} \left( \exp\left[\frac{\Delta\Gamma^{\text{DL}}}{\Gamma^{\text{Born}}}\right]
  - \left[1
  + \frac{\Delta\Gamma^{\text{DL}}}{\Gamma^{\text{Born}}}\right] \right)\!\right\}
\end{align}
where~$\Gamma^{\text{Born}}$, $\Gamma^{\text{eff}}$,
$\Gamma^{\text{QCD+QED}}$ and~$\Gamma^{\text{FO}}$ have already been
defined above. The shift~$\Delta\Gamma^{\text{DL}}$ is obtained
directly from the width~$\Gamma^{\text{FO}}$ that is derived from the
automated calculation at fixed order by substituting the loop
functions with the associated double~logarithms. In
Eq.\,\eqref{eq:impwidth}, the two subtracted terms in the square
brackets avoid double~counting between the `tree-level improved'
width~$\Gamma^{\text{eff}}\,\exp{\left(\Delta\Gamma^{\text{DL}}/\Gamma^{\text{Born}}\right)}$
and the full fixed-order one-loop width~$\Gamma^{\text{FO}}$.

\newpage

\tocsubsection{Real radiation}

\subsubsection*{Full calculation}

From the perspective of order-counting, the three-body decay widths at
the tree level are of the same order as one-loop corrections to the
two-body decays. For a full one-loop evaluation of the total widths
(and branching ratios) of the Higgs bosons, it is thus mandatory to
include these channels in the analysis. If we restrict ourselves to
non-SUSY~final states, the only possible relevant three-body final
states involve the radiation of gauge bosons or light Higgs
bosons. The radiation of photons and gluons was actually already taken
into account above, hence included within the two-body decay widths.

Using \texttt{FeynArts} and \texttt{FormCalc}, we perform the
calculation of the widths at the tree-level order for the following
channels: \AtoB{h_i}{Zf\bar{f}}, \AtoB{H^{\pm}}{W^{\pm}f\bar{f}},
\AtoB{h_i}{h_jf\bar{f}} for~\mbox{$f\in\{b,t,\tau\}$}, and
\AtoB{h_i}{W^-f\bar{f'}}, \AtoB{h_i}{W^+f'\bar{f}},
\AtoB{h_i}{H^-f\bar{f'}}, \AtoB{h_i}{H^+f'\bar{f}},
\AtoB{H^+}{Zf\bar{f'}}, \AtoB{H^+}{h_jf\bar{f'}}
for~\mbox{$(f,f')\in\{(t,b),(\tau^+,\nu_{\tau})\}$}. We stress that
contrarily to the radiation of photons and gluons, these final states
are clearly distinguishable from the fermionic two-body channels, so
that there is no deep reason to consider (only) inclusive decay widths
from a theoretical perspective.

The kinematical integral can be performed (\EG) in the rest frame of
the fermion pair in the final state. There, only two steps of
integration are non-trivial, one applying to an `angular' variable,
the other to the invariant mass-squared~$s$ of the fermion pair. We
perform the angular integral analytically, while that on~$s$ is
carried out numerically. The internal lines are systematically
regularized by inserting the total width~$\Gamma$ of the corresponding
particles in the propagators. These `internal' widths are calculated
from the sum of all relevant two-body processes at the tree level.

Having set up the calculation in this way, the leading terms in the
squared amplitudes scale
like~\mbox{$\Gamma{\left[\AtoB{h}{Vf\bar{f}'}\right]}/\Gamma{\left[\AtoB{h}{f\bar{f}'}\right]}\propto
M_h^2/M_V^2$}. This would lead to the unphysical situation where the
three-body decay widths grow faster than the two-body widths, and
eventually dominate the decays. However, these contributions linear
in~$M_h^2/M_V^2$ sum up to zero as long as the heavy Higgs doublets
decouple from the electroweak sector. For the sake of not spoiling the
physical interpretation, it is thus paramount to preserve these
decoupling properties of the heavy Higgs states at the level of the
loop-corrected Higgs-mixing matrix.

\subsubsection*{On-shell contributions}

When an internal line can be exchanged on-shell, the corresponding
contribution near the pole (after dismissing the quadratically growing
terms discussed above) is dominated by the narrow-width
approximation. In this case, a piece of the three-body width is
already accounted for at the level of the two-body widths. Therefore,
we subtract such on-shell contributions explicitly. This step is
performed by evaluating the two-body widths and branching ratios that
intervene in the considered process at the tree level. It is crucial
that the full widths that are employed as regulators in the internal
lines are computed at exactly the same order as these intermediate
two-body widths.

We note that the narrow-width approximation tends to overestimate the
contribution from the on-shell pole~$M$: this can be easily understood
by comparing the lorentzian distribution and its limit
for~\mbox{$\Gamma\to 0$},
\begin{align}
  \left[(s - M^2)^2 + M^2\,\Gamma^2\right]^{-1} &\xrightarrow{\Gamma\to 0}
  \frac{\pi}{M\,\Gamma}\,\delta{\left(s-M^2\right)}\,.
\end{align}
Consequently, the subtraction of the on-shell contributions can result
in an apparent negative width (as long as the Sudakov terms remain
subleading), which, however, should be interpreted as a correction
applying to the two-body width.

In addition, we expect the radiation of electroweak gauge~bosons and
Higgs bosons to produce Sudakov double and single~logarithms again, when
compared to the corresponding two-body width at the~Born~level. It is
useful to extract them explicitly.

\subsubsection*{Double logarithms}

\begin{figure}[b!]
  \centering
  \includegraphics[width=.7\textwidth]{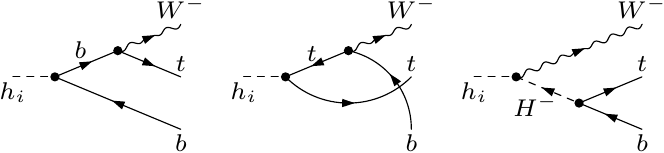}
  \caption{\label{fig:sud_radiation} Contributions
  to~\AtoB{h_i}{W^-t\bar{b}} leading to double logarithms.}
\end{figure}

The double~logarithms are associated with the radiation of electroweak
gauge~bosons. The topologies leading to Sudakov~logarithms are
summarized in Fig.\,\ref{fig:sud_radiation} for the example
of~\AtoB{h_i}{W^-t\bar{b}}. The topology with an internal Higgs line
is relevant only if the external and internal states are close in
mass. For heavy doublet-like Higgs bosons in the initial state, we
obtain:
\begin{subequations}
\begin{align}
  \GamRat{DL}{\AtoB{h_i}{Zb\bar{b}\,}}{}{\AtoB{h_i}{b\bar{b}\,}}{} &\simeq
  \frac{1}{48\,\pi^2}\left[
  \frac{5}{12}\,g_1^2 + \frac{3}{4}\,g_2^2 - \frac{2}{3}\,e^2\right]
  \ln^2\frac{M_V^2}{M_{h_i}^2}\,,\\
  \GamRat{DL}{\AtoB{h_i}{Zt\bar{t}\,}}{}{\AtoB{h_i}{t\bar{t}\,}}{} &\simeq
  \frac{1}{48\,\pi^2}\left[
  \frac{17}{12}\,g_1^2 + \frac{3}{4}\,g_2^2 - \frac{8}{3}\,e^2\right]
  \ln^2\frac{M_V^2}{M_{h_i}^2}\,,\\
  \GamRat{DL}{\AtoB{h_i}{Z\tau^+\tau^-}}{}{\AtoB{h_i}{\tau^+\tau^-}}{}
  &\simeq \frac{1}{48\,\pi^2}\left[
  \frac{15}{4}\,g_1^2 + \frac{3}{4}\,g_2^2 - 6\,e^2\right]
  \ln^2\frac{M_V^2}{M_{h_i}^2}\,,\\
  \GamRat{DL}{\AtoB{h_i}{W^-t\bar{b}\,}}{b}{\AtoB{h_i}{b\bar{b}\,}}{} &\simeq
  \GamRat{DL}{\AtoB{h_i}{W^-t\bar{b}\,}}{t}{\AtoB{h_i}{t\bar{t}\,}}{}\simeq
  \GamRat{DL}{\AtoB{h_i}{W^-\tau^+\nu_{\tau}}}{}{\AtoB{h_i}{\tau^+\tau^-}}{}
  \simeq \frac{g_2^2}{64\,\pi^2}\ln^2\frac{M_V^2}{M_{h_i}^2}\,,\\
  \GamRat{DL}{\AtoB{H^-}{Zb\bar{t}\,}}{b}{\AtoB{H^-}{b\bar{t}\,}}{b} &\simeq
  \frac{1}{48\,\pi^2}\left[
  \frac{5}{12}\,g_1^2 + \frac{3}{4}\,g_2^2 - \frac{5}{3}\,e^2\right]
  \ln^2\frac{M_V^2}{M_{H^{\pm}}^2}\,,\\
  \GamRat{DL}{\AtoB{H^-}{Zb\bar{t}\,}}{t}{\AtoB{H^-}{b\bar{t}\,}}{t} &\simeq
  \frac{1}{48\,\pi^2}\left[
  \frac{17}{12}\,g_1^2 + \frac{3}{4}\,g_2^2 - \frac{5}{3}\,e^2\right]
  \ln^2\frac{M_V^2}{M_{H^{\pm}}^2}\,,\\
  \GamRat{DL}{\AtoB{H^-}{Z\tau^-\bar{\nu}_{\tau}}}{}{\AtoB{H^-}{\tau^-\bar{\nu}_{\tau}}}{}
  &\simeq \frac{1}{48\,\pi^2}\left[
  \frac{15}{4}\,g_1^2 + \frac{3}{4}\,g_2^2 - 3\,e^2\right]
  \ln^2\frac{M_V^2}{M_{H^{\pm}}^2}\,,\\
  \GamRat{DL}{\AtoB{H^-}{W^-b\bar{b}\,}}{b}{\AtoB{H^-}{b\bar{t}\,}}{b} &\simeq
  \GamRat{DL}{\AtoB{H^-}{W^-t\bar{t}\,}}{t}{\AtoB{H^-}{b\bar{t}\,}}{t}\simeq
  \GamRat{DL}{\AtoB{H^-}{W^-\tau^+\tau^-}}{}{\AtoB{H^-}{\tau^-\bar{\nu}_{\tau}}}{}
  \simeq\frac{g_2^2}{32\,\pi^2} \ln^2\frac{M_V^2}{M_{H^{\pm}}^2}\,.
\end{align}
\end{subequations}
As expected, these double logarithms exactly cancel with those of
Eq.\,\eqref{eq:2bDL} for a given initial state. This means that the
sizable shift associated with double~logarithms affects the exclusive
two-body decay widths, but not the inclusive one (including radiation
of~$W$ and~$Z$) or the total one.

If we adopt the perspective of a resummation of Sudakov
double~logarithms beyond three-particle final states, we can define
the electroweak-radiation width~$\Gamma^{\text{EW rad}}$ for a very
massive Higgs boson in the initial state in terms of the quantities
intervening in the two-body decays:
\begin{equation}
  \frac{\Gamma^{\text{EW rad}}}{\Gamma^{\text{Born}}}=
  1-\exp\left[\frac{\Delta\Gamma^{\text{DL}}}{\Gamma^{\text{Born}}}\right].
\end{equation}
This object resums the radiation of~$W$ and~$Z$~bosons, as well as
additional photon emission at higher orders related to the mixing of
abelian and non-abelian parts in the electroweak gauge group
(see \citere{Fadin:1999bq}), that is associated with the two-body
width~$\Gamma^{\text{Born}}$.

\subsubsection*{Single logarithms}

The single logarithms originate in electroweak radiation processes but
also in Higgs-radiation processes. For the logarithms of electroweak
type, we find the following contributions to the decays of
doublet-like states:
\begin{subequations}\label{eq:3bSL}
\begin{align}
  \GamRat{SL}{\AtoB{h_i}{Zb\bar{b}\,}}{}{\AtoB{h_i}{b\bar{b}\,}}{} &\simeq
  \frac{1}{16\,\pi^2}\left[
  \frac{11}{12}\,g_1^2 + \frac{5}{4}\,g_2^2 - \frac{2}{3}\,e^2\right]
  \ln\frac{M_V^2}{M_{h_i}^2}\,,\\
  \GamRat{SL}{\AtoB{h_i}{Zt\bar{t}\,}}{}{\AtoB{h_i}{t\bar{t}\,}}{} &\simeq
  \frac{1}{16\,\pi^2}\left[
  \frac{23}{12}\,g_1^2 + \frac{5}{4}\,g_2^2 - \frac{8}{3}\,e^2\right]
  \ln\frac{M_V^2}{M_{h_i}^2}\,,\\
  \GamRat{SL}{\AtoB{h_i}{Z\tau^+\tau^-}}{}{\AtoB{h_i}{\tau^+\tau^-}}{}
  &\simeq \frac{1}{16\,\pi^2}\left[
  \frac{17}{4}\,g_1^2 + \frac{5}{4}\,g_2^2 - 6\,e^2\right]
  \ln^2\frac{M_V^2}{M_{h_i}^2}\,,\\
  \GamRat{SL}{\AtoB{h_i}{W^-t\bar{b}\,}}{b}{\AtoB{h_i}{b\bar{b}\,}}{} &\simeq
  \GamRat{SL}{\AtoB{h_i}{W^-t\bar{b}\,}}{t}{\AtoB{h_i}{t\bar{t}\,}}{}\simeq
  \GamRat{SL}{\AtoB{h_i}{W^-\tau^+\nu_{\tau}}}{}{\AtoB{h_i}{\tau^+\tau^-}}{}
  \simeq\frac{5\,g_2^2}{64\,\pi^2}\ln\frac{M_V^2}{M_{h_i}^2}\,,\\
  \GamRat{SL}{\AtoB{H^-}{Zb\bar{t}\,}}{b}{\AtoB{H^-}{b\bar{t}\,}}{b} &\simeq
  \frac{1}{16\,\pi^2}\left[
  \frac{11}{12}\,g_1^2 + \frac{5}{4}\,g_2^2 - \frac{11}{3}\,e^2\right]
  \ln\frac{M_V^2}{M_{H^{\pm}}^2}\,,\\
  \GamRat{SL}{\AtoB{H^-}{Zb\bar{t}\,}}{t}{\AtoB{H^-}{b\bar{t}\,}}{t}&\simeq
  \frac{1}{16\,\pi^2}\left[
  \frac{23}{12}\,g_1^2 + \frac{5}{4}\,g_2^2 - \frac{11}{3}\,e^2\right]
  \ln\frac{M_V^2}{M_{H^{\pm}}^2}\,,\\
  \GamRat{SL}{\AtoB{H^-}{Z\tau^-\bar{\nu}_{\tau}}}{}{\AtoB{H^-}{\tau^-\bar{\nu}_{\tau}}}{}
  &\simeq\frac{1}{16\,\pi^2}\left[
  \frac{17}{4}\,g_1^2 + \frac{5}{4}\,g_2^2 - 5\,e^2\right]
  \ln\frac{M_V^2}{M_{H^{\pm}}^2}\,,\\
  \GamRat{SL}{\AtoB{H^-}{W^-b\bar{b}\,}}{b}{\AtoB{H^-}{b\bar{t}\,}}{b} &\simeq
  \GamRat{SL}{\AtoB{H^-}{W^-t\bar{t}\,}}{t}{\AtoB{H^-}{b\bar{t}\,}}{t}\simeq
  \GamRat{SL}{\AtoB{H^-}{W^-\tau^+\tau^-}}{}{\AtoB{H^-}{\tau^-\bar{\nu}_{\tau}}}{}
  \simeq\frac{5\,g_2^2}{32\,\pi^2}\ln\frac{M_V^2}{M_{H^{\pm}}^2}\,.
\end{align}
\end{subequations}
We note that, for a given initial state, if we add the electroweak
single logarithms arising in the two-body decays given in
Eqs.\,\eqref{eq:2bSL} (amounting to zero) and those appearing in the
three-body decays in Eqs.\,\eqref{eq:3bSL}, we recover the electroweak
logarithms expected from the one-loop~RGEs of the Yukawa couplings, up
to two pieces:
\begin{itemize}
\item
  the logarithms associated to QED~contributions, corresponding to the
  terms in~$e^2$, are processed in the~QED-correction factor and are
  thus subtracted from the
  full electroweak contribution;
\item
  the logarithms associated to the Higgs wave~function exactly cancel
  the electroweak single~logarithms from the vertex corrections to the
  two-body decays: we recover the logarithms of the~RGEs only if we
  subtract the logarithms from the Higgs field normalization (the
  latter correspond to the~RGEs of the Higgs fields).
\end{itemize}
This result is physically intuitive: as in the case of
QCD~corrections, only logarithms of UV~type remain at the level of the
inclusive width, hence matching what is expected from the anomalous
dimension of the operator mediating the decay. It would thus be
possible to absorb these logarithms in a version of the Yukawa
couplings including the electroweak running. However, this would hold
only at the level of the inclusive width, while for the exclusive
widths individual logarithms of IR~type would have to be subtracted
again.

In order to count the logarithms associated to the emission of light
Higgs bosons in the decays of a heavy Higgs state, we can consider the
radiation of a SM-like Higgs boson and that of a light
singlet-dominated state separately. The radiation of a heavy
doublet-like state is kinematically forbidden (or suppressed) in
general: it would only be relevant for the decays of an even heavier
singlet state. For the radiation of a SM-like Higgs boson from a
heavy-doublet initial state, the relevant topologies are those where
the light Higgs emerges from a fermion line. Indeed, the topologies
with a triple-Higgs or a Higgs--vector coupling are always suppressed,
because either the coupling itself leads to a suppression factor of
electroweak size, or the mixing angles vanish due to the decoupling
limit, or the couplings of a singlet-dominated scalar with SM~fermions
are small. Then, at the numerical level, the radiation of a SM-like
Higgs is mostly relevant in association with a top-quark line, since
the coupling to lighter fermions is suppressed in proportion with
their masses. Correspondingly, we obtain the following leading
logarithms:
\begin{subequations}
\begin{align}
  \GamRat{SL}{\AtoB{h_i}{h_{\text{SM}}t\bar{t}\,}}{h}{\AtoB{h_i}{t\bar{t}\,}}{}
  &\simeq -\frac{Y_t^2\,s_{\beta}^2}{32\,\pi^2}
  \ln\frac{M_{h_{\text{SM}}}^2}{M_{h_i}^2}\,,\\
  \GamRat{SL}{\AtoB{H^-}{h_{\text{SM}}b\bar{t}\,}}{h}{\AtoB{H^-}{b\bar{t}\,}}{}
  &\simeq -\frac{Y_t^2\,s_{\beta}^2}{64\,\pi^2}
  \ln\frac{M_{h_{\text{SM}}}^2}{M_{H^{\pm}}^2}\,.
\end{align}
\end{subequations}
Additional logarithms of Yukawa type also appear in the radiation of
weak gauge~bosons, replacing the contributions by radiated
Goldtone~bosons in the limit of conserved electroweak symmetry.

The radiation of a light singlet by a heavy doublet state is dominated
by the topology with a triple-Higgs coupling, with the internal line
possessing a mass close to that of the decaying state. The couplings
between heavy doublet~states and a singlet are typically of the order
of~$\sqrt{2}\,\lambda\,\mu_{\text{eff}}$. Thus the radiation of a
singlet-like Higgs provides an unsuppressed logarithm if this coupling
fulfills the
condition~\mbox{$\sqrt{2}\,\lambda\,\mu_{\text{eff}}=\mathcal{O}{\left(M_{H^{\pm}}\right)}$}.

\subsubsection*{Inclusive three-body width}

Our automated calculation provides us with each individual three-body
width~$\Gamma^{\text{3b}}$, from which the on-shell contributions have
been subtracted. We then define a total width for all the three-body
decays. In order to resum the double logarithms consistently with
Eq.\,\eqref{eq:impwidth}, we schematically define the inclusive
three-body width as
\begin{align}\label{eq:res3bwidth}
  \Gamma^{\text{3b, incl}} &= \sum_{\text{2b}}
  \Gamma^{\text{eff}}\cdot\frac{\Gamma^{\text{QCD+QED}}}{\Gamma^{\text{Born}}}\left\{
    1 - \exp\left[\frac{\Delta\Gamma^{\text{DL}}}{\Gamma^{\text{Born}}}\right]
    + \frac{\Delta\Gamma^{\text{DL}}}{\Gamma^{\text{Born}}}\right\}
  + \sum_{\text{3b}}\Gamma^{\text{3b}}
\end{align}
where the last term inside the curly brackets avoids double counting
between the resummed piece (first two terms inside the curly brackets)
and the diagrammatic calculation (last term). In
calculating~$\Gamma^{\text{3b}}$, we include a rescaling of the widths
following the~QCD and QED~corrections of the associated two-body
decay. In particular Yukawa couplings are run to the scale of the
decaying Higgs. In addition, we also take the short-distance effect of
SUSY~particles into account. These corrections are formally of higher
order but have a sizable numerical impact. In particular, the use of
`naive' Yukawa couplings would make the three-body widths almost
systematically larger than their two-body counterparts. The
correspondence of electroweak logarithms on both sides, the two-body
and three-body decays, serves us as a guideline as to the proper
definition of the tree-level couplings employed in the three-body
calculation.

The inclusive width for electroweak radiation defined in
Eq.\,\eqref{eq:res3bwidth} relies on several assumptions that restrict
its validity. In particular, the kinematics does not allow for the
radiation of an infinite number of weak gauge bosons, meaning that the
exponential series of double-logarithmic terms is in fact explicitly
truncated. Yet, in the considered regime the mismatch between the
truncated series and the exponential is negligible. As we will show in
the following section, the double-logarithmic effects amount
to~$\simord10\%$ for states at~$1$\,TeV and~$\simord40\%$
at~$100$\,TeV: comparing~$\exp{(0.1)}$ and~$\exp{(0.4)}$ with their
Taylor expansion at~$0$, we observe that the difference remains at or
below the permil level already after including the third order,
whereas the kinematics allows for a much larger number of final states
with electroweak mass. The description of radiated gauge bosons as
actual final states forms another limitation, as~$W$ and~$Z$ in fact
decay into fermions. The `genuine' process is therefore more complex
and includes further topologies, beyond those obtained from
radiated~$W$ and~$Z$~bosons. We believe, however, that this simplified
representation still captures the leading contributions to the Higgs
decays beyond two-body final states, since, in the considered regime,
we do not expect larger effects than the double logarithms, the latter
being intimately associated with gauge-boson radiation.

%% file: 03_NumericalAnalysis.tex
\tocsection[\label{sec:numerics}]{Numerical Analysis}

In this section, we present the consequences of the electroweak
corrections to the fermionic decays at the numerical level. Below, we
focus on the heavy doublet states in a scenario with decoupled
SUSY~sector (masses of the order of~$100$\,TeV). The
parameter~$\tan\beta$ is kept at the intermediate value~$10$
while~\mbox{$\lambda=\kappa\ll1$}. We then vary the mass of the
charged Higgs, also controlling that of the heavy \cp-even and \cp-odd
doublet states~$h_2$ and~$h_4$, between~$1$\,TeV and~$100$\,TeV. While
the physical relevance of this scenario---\EG\ with respect to the
Hierarchy Problem---can be questioned, it is entirely meant to clearly
exemplify the impact of the electroweak corrections in extensions of the SM based on a THDM structure.

\begin{figure}[tbp!]
  \centering
  \includegraphics[width=0.48\textwidth]{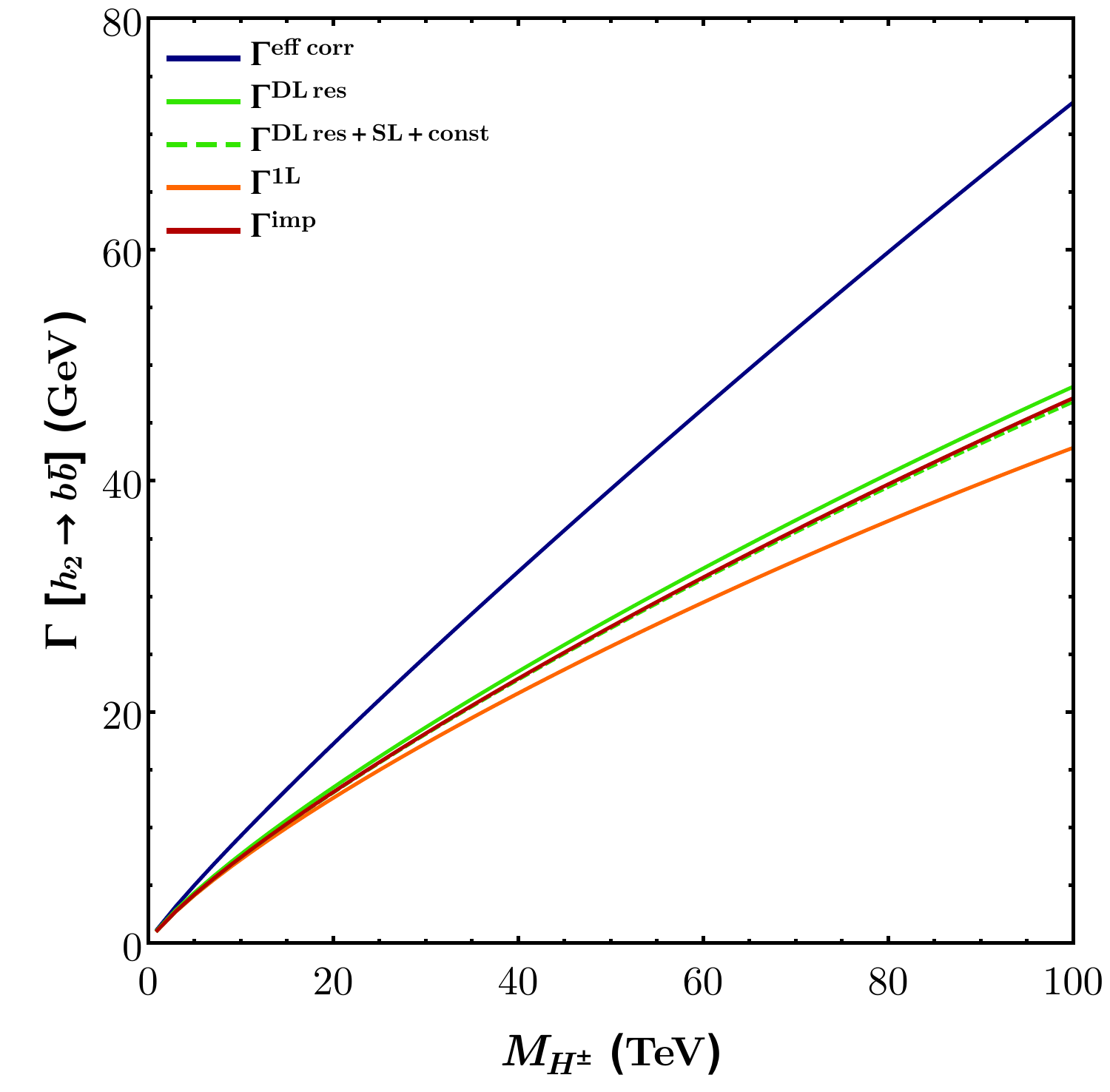}
  \includegraphics[width=0.48\textwidth]{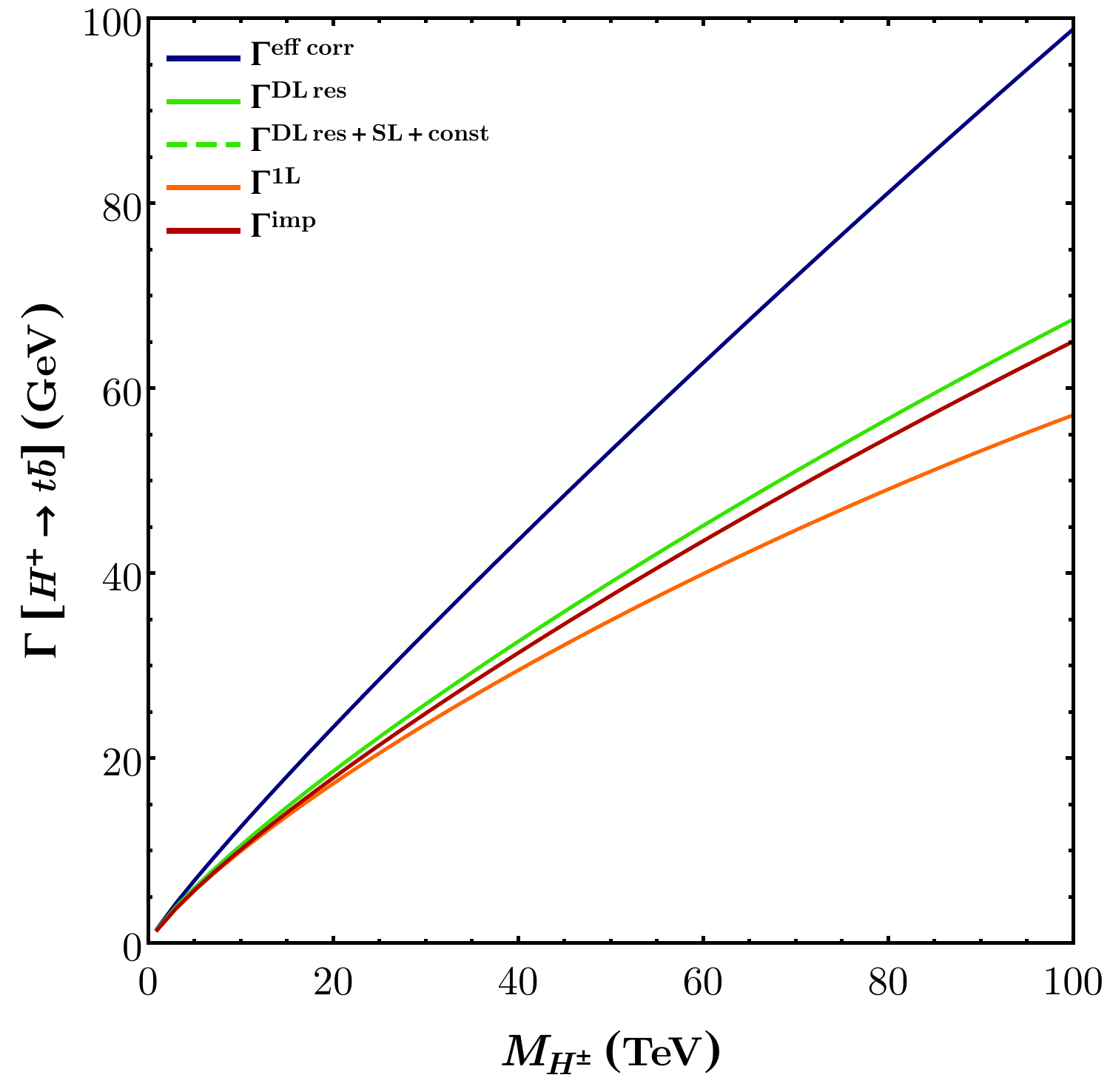}\\
  \includegraphics[width=0.48\textwidth]{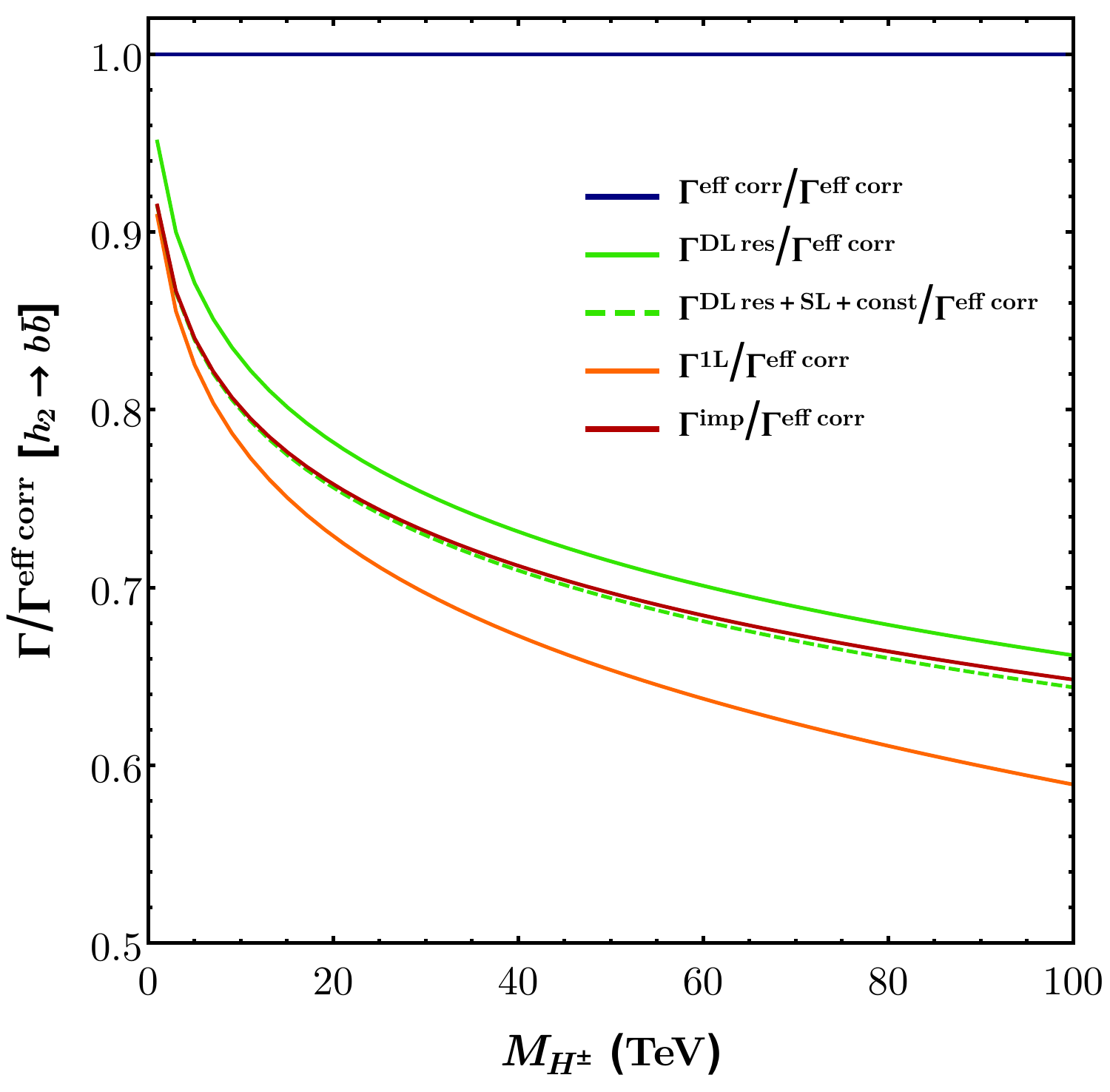}
  \includegraphics[width=0.48\textwidth]{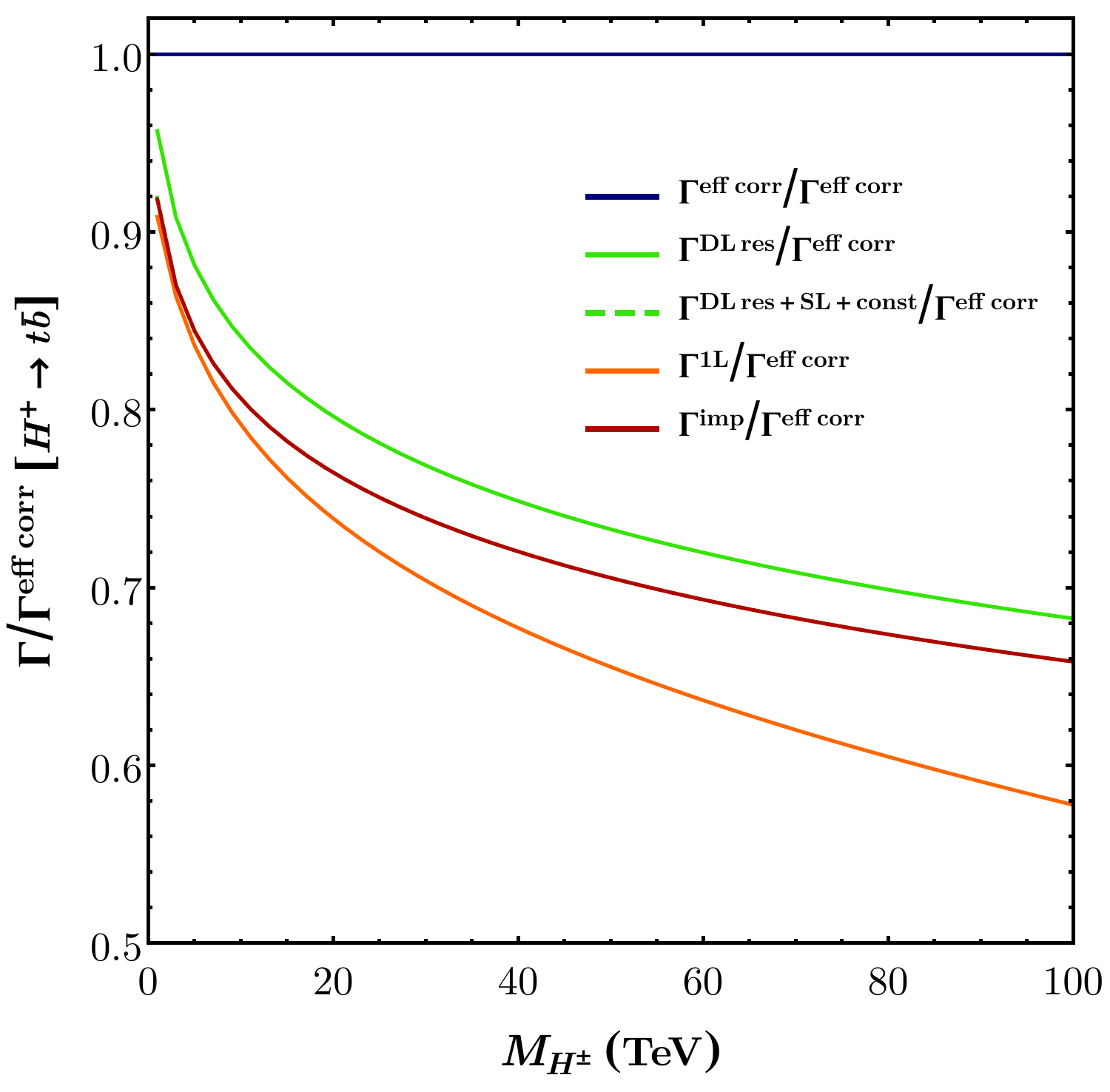}
  \caption{\label{fig:hbb_width}
  The electroweak corrections to the Higgs decay width
  into~$b\bar{b}$~final states are shown for a~\cp-even heavy doublet
  state (left-hand column). The corresponding plots for the~\cp-odd
  doublet state are virtually identical. The plots in the column on
  the right-hand side correspond to the charged-Higgs
  decay~\AtoB{H^+}{t\bar{b}}. We display the decay widths in the first
  row, whereas they are normalized to~$\Gamma^{\text{eff}}$ in the
  second row for a better appreciation of the magnitude of the
  effects. The blue~curve represents the prediction
  for~$\Gamma^{\text{eff}}$, the effective tree-level width including
  short-distance corrections of SUSY~particles as well as
  long-distance effects from~QCD and~QED. The green~solid~curve
  additionally accounts for the Sudakov double~logarithms while the
  green~dashed~curve also evaluates the impact of single~logarithms
  and constant terms. The orange~curve is the full one-loop evaluation
  at fixed order obtained from the diagrammatic calculation. For the
  red~curve we combine a resummation of the double~logarithms with
  this full one-loop evaluation, according to
  Eq.\,\eqref{eq:impwidth}. In the case of~\AtoB{H^+}{t\bar{b}}, the
  green~dashed~curve is hidden by the red~one.}
\end{figure}

\subsubsection*{Two-body decay width}

We first consider the decay width into final states including
(anti)quarks of the third generation for heavy doublet-like states in
the scenario presented above. In Fig.\,\ref{fig:hbb_width}, we plot
these decay widths against the mass of the charged Higgs~$M_{H^{\pm}}$
(we recall that~\mbox{$M_{H^{\pm}}\approx M_{h_2}\approx M_{h_4}$}) in
several approximations. The~QCD and QED~corrections are included per
default, as well as the decoupling SUSY~effects. For the blue~curve,
we only consider the width resulting from this level of approximation
(denoted as~\mbox{$\Gamma^{\text{eff
corr}}=\Gamma^{\text{eff}}\,\Gamma^{\text{QCD+QED}}/\Gamma^{\text{Born}}$}
in the following): this is essentially the popular description
employed in public codes such as~\texttt{HDECAY}. The
green~solid~curve corresponds to the `improved~tree-level' result,
where Sudakov double~logarithms are included:
\begin{align}
  \Gamma^{\text{DL res}} &= \frac{\Gamma^{\text{QCD+QED}}}{\Gamma^{\text{Born}}}\,
  \Gamma^{\text{eff}}\,\exp{\left(
  \frac{\Delta\Gamma^{\text{DL}}}{\Gamma^{\text{Born}}}\right)}\,.
\end{align}
The green~dashed~curve is obtained after additionally incorporating
the single~logarithms and constant (non-logarithmic) pieces. The
result of the full diagrammatic one-loop calculation
of \citere{Domingo:2018uim} (denoted
as~\mbox{$\Gamma^{\text{1L}}=\Gamma^{\text{FO}}\,\Gamma^{\text{QCD+QED}}/\Gamma^{\text{Born}}$}
in the following) is displayed in orange. Finally, we show the full
one-loop width including resummation of the Sudakov double~logarithms
($\Gamma^{\text{imp}}$ of Eq.\,\eqref{eq:impwidth}) in red. In the
second row, we show the same estimates normalized
to~$\Gamma^{\text{eff}}$. The left-hand side focuses on the decay
channel~\AtoB{h_2}{b\bar{b}} of the \cp-even doublet-like state~$h_2$,
while the column on the right-hand side
considers~\AtoB{H^+}{t\bar{b}}. The effects are largely comparable in
both cases.

As expected, the impact of the electroweak corrections
reaches~$\mathcal{O}{\left(10\%\right)}$
for~\mbox{$M_{H^{\pm}}\sim1$}\,TeV and more than~$40\%$
at~\mbox{$M_{H^{\pm}}\sim100$}\,TeV. While the tree-level-improved
width, including the resummed double~logarithms (green~solid~curve),
gives a good qualitative approximation, it is still~$\simord5\%$~off
as compared to the full result. The agreement can be further refined
if we account for the single~logarithms and constant terms
(green~dashed~curve; considering only the single logarithms would lead
to an approximation of intermediate quality). Obviously, the
resummation of Sudakov double~logarithms only yields an appreciable
effect beyond~\mbox{$M_{H^{\pm}}\gsim10$}\,TeV, as the red and
orange~curves remain comparatively close below this mark. It only
amounts to~$\mathcal{O}{\left(5\text{--}10\%\right)}$ as compared to
$\Gamma^{\text{1L}}$ beyond.

\subsubsection*{Three-body widths}

\begin{figure}[tbp!]
  \centering
  \includegraphics[width=0.48\textwidth]{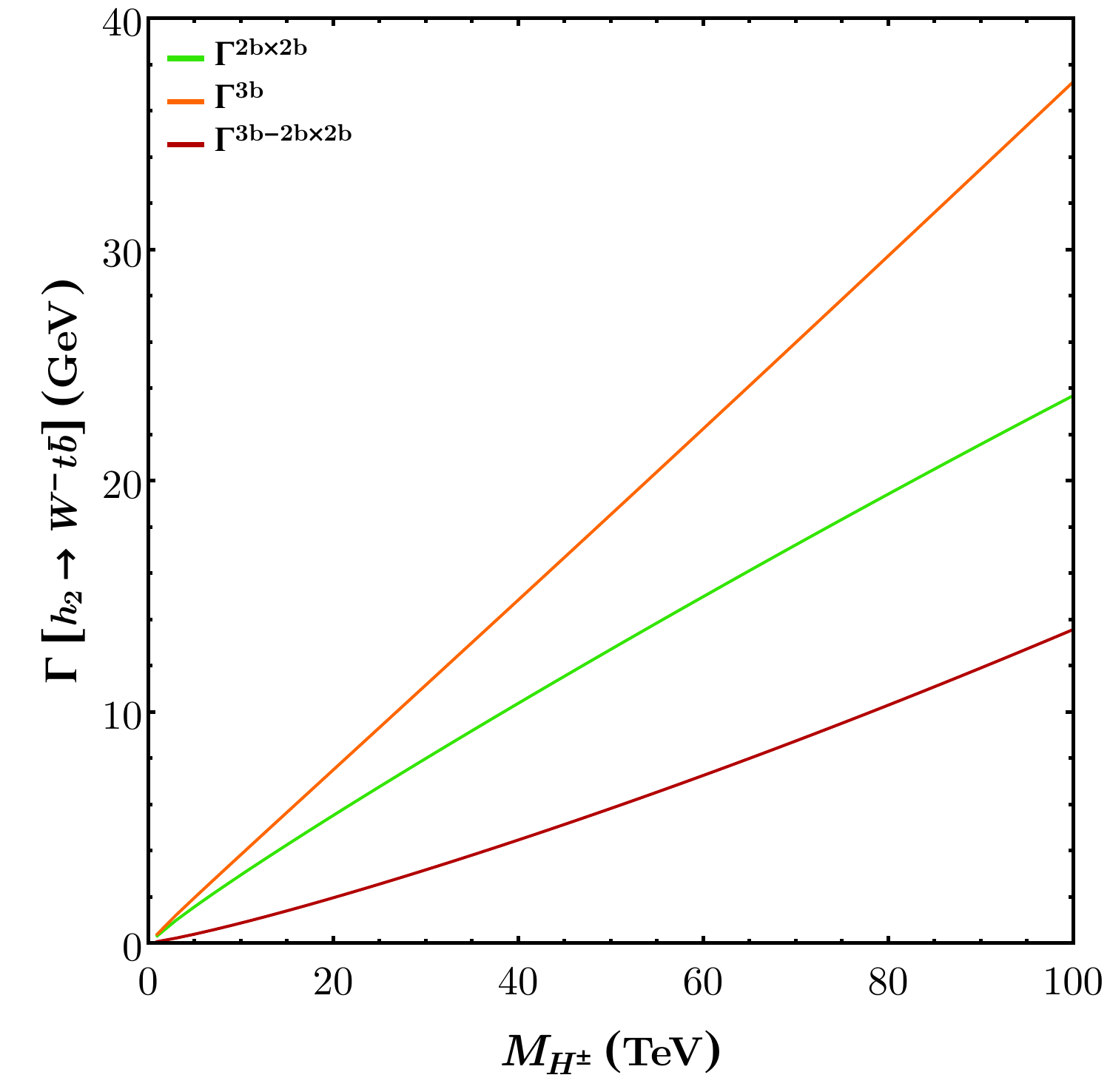}
  \includegraphics[width=0.48\textwidth]{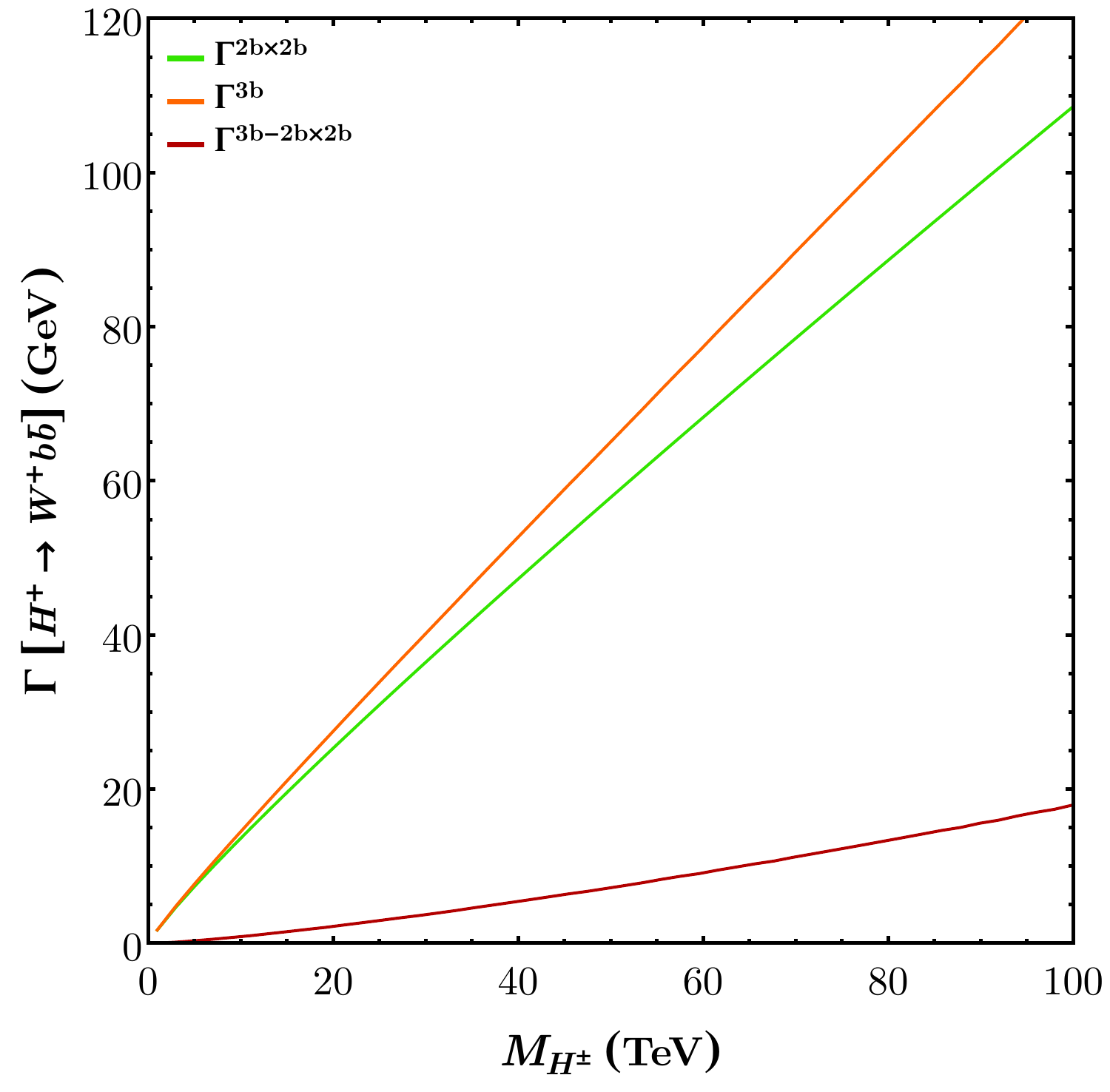}\\
  \includegraphics[width=0.48\textwidth]{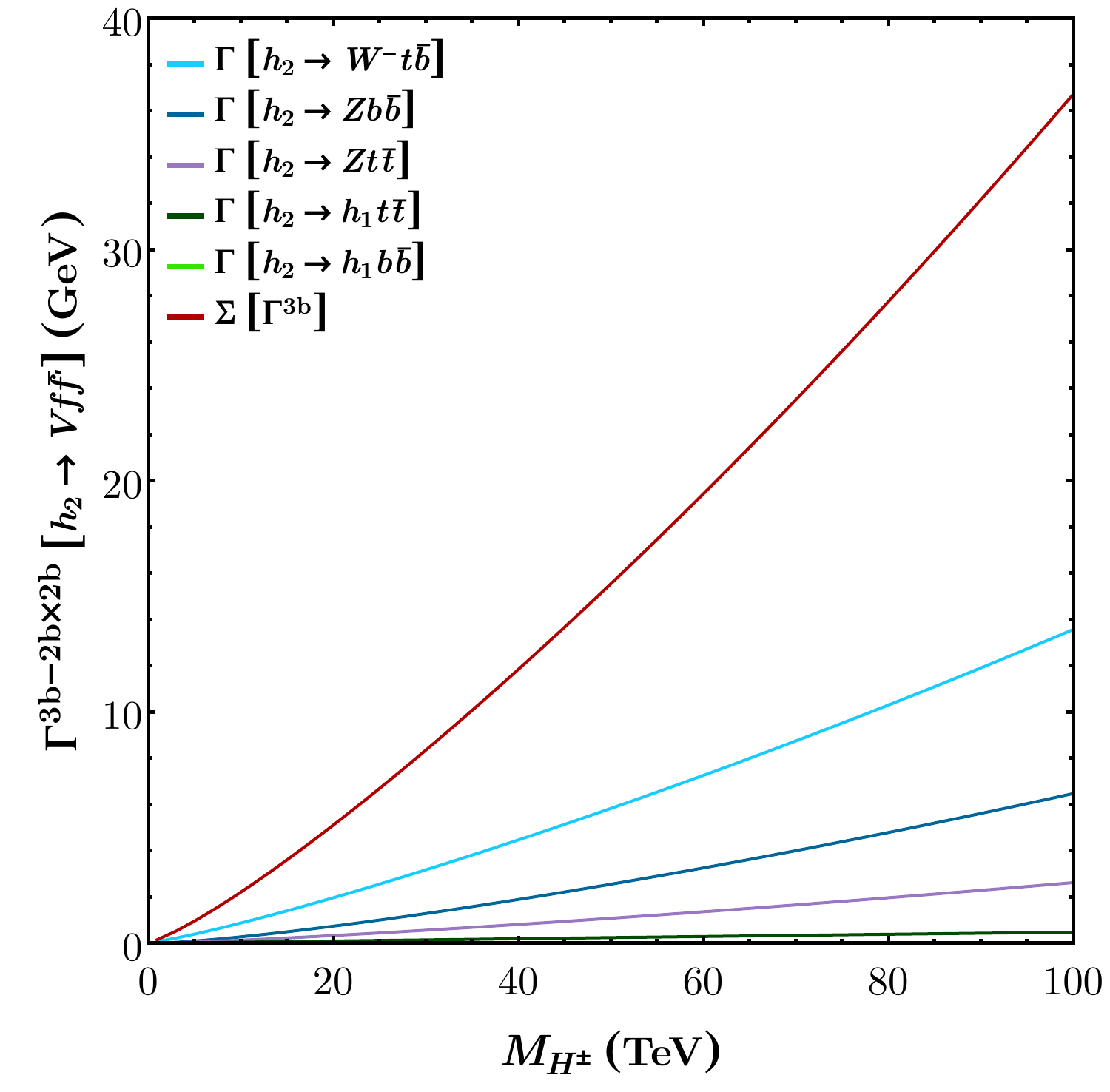}
  \includegraphics[width=0.48\textwidth]{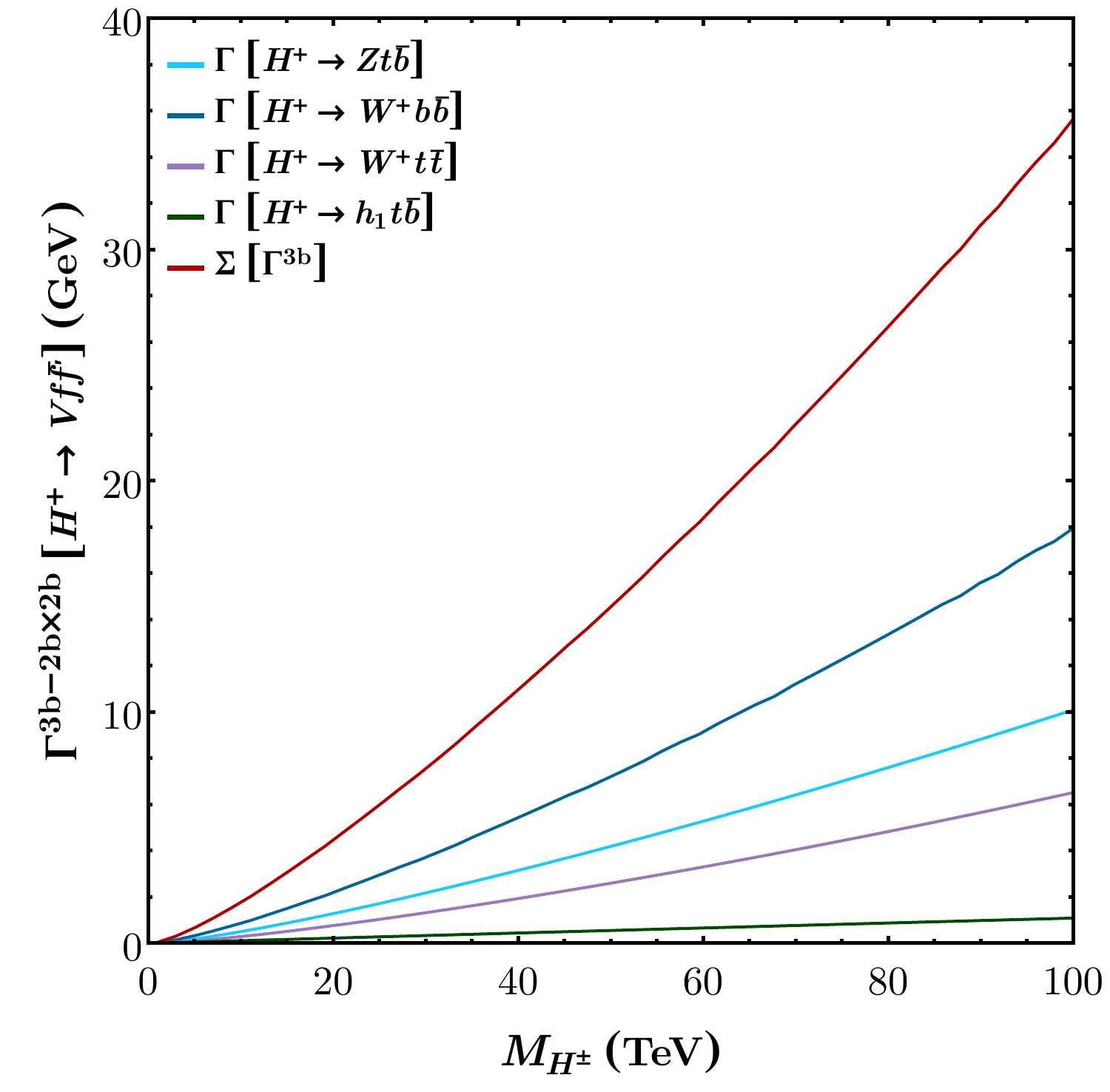}
  \caption{\label{fig:3b_width}
  The fermionic three-body decay widths of a heavy decoupling~\cp-even
  doublet (left column) and a charged Higgs (right
  column) are shown.
  \newline\textit{Top}: the decay widths for the
  channels~\AtoB{h_2}{W^-t\bar{b}} (left) and~\AtoB{H^+}{W^+b\bar{b}}
  (right); we display the full three-body width in orange; in green,
  we show the two-body contributions mediated by an intermediate
  on-shell line (in
  particular~\AtoB{h_2}{t(\AtoB{\bar{t}}{W^-\bar{b}})}
  and~\AtoB{H^+}{(\AtoB{t}{W^+b})\bar{b}}); the red~curve corresponds
  to the genuine three-body contribution beyond the two-body
  approximation (difference of the orange and green~curves).
  \newline\textit{Bottom}: the various three-body contributions beyond
  the two-body approximation, considering only top and bottom~quarks
  in the final state; the various channels with radiation of a gauge
  boson are displayed in shades of blue; radiation of a SM-like Higgs
  boson is shown in shades of green; the red curve corresponds to the
  sum of all these individual channels.  }
\end{figure}

Now, we turn to the three-body decays. In the upper row of
Fig.\,\ref{fig:3b_width}, we show the widths of the three-body
channels~\AtoB{h_2}{W^-t\bar{b}}~(left)
and~\AtoB{H^+}{W^+b\bar{b}}~(right) in the same scenario as in
Fig.\,\ref{fig:hbb_width}. The orange~curve corresponds to the full
three-body width, while the contribution from on-shell intermediate
particles (in particular~\AtoB{t}{W^+b}) is shown in green. The
resulting genuine three-body contribution (beyond the two-body
approximation) is shown in red and corresponds to the piece dominated
by Sudakov double~logarithms. We observe that this off-shell component
(we continue to use this terminology below) is sizable and of
competitive magnitude as compared to the two-body widths.

In the lower row of Fig.\,\ref{fig:3b_width}, we display the off-shell
component of the three-body widths for the various relevant three-body
channels (restricting ourselves to top and bottom~quarks in the final
state). Their sum is displayed in red. We observe that these
contributions roughly reach the magnitude of the electroweak effects
in the two-body width, corresponding to the difference between the
blue and red~curves in Fig.\,\ref{fig:hbb_width}. We also stress that
the three-body contribution is not of negligible size as compared to
the two-body widths, so that we expect a sizable impact at the level
of the branching ratios.

\begin{figure}[tbp!]
  \centering
  \includegraphics[width=0.48\textwidth]{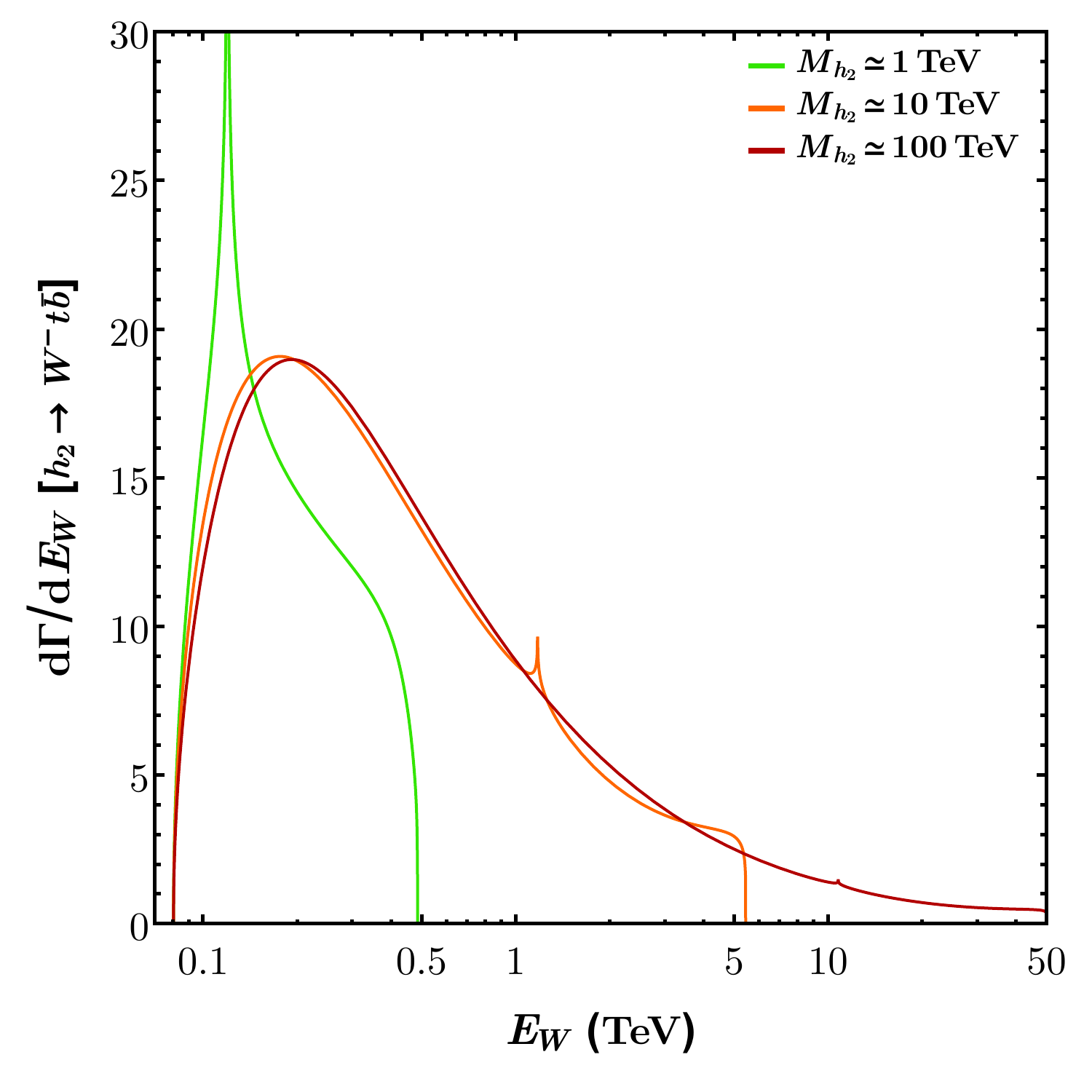}
  \includegraphics[width=0.48\textwidth]{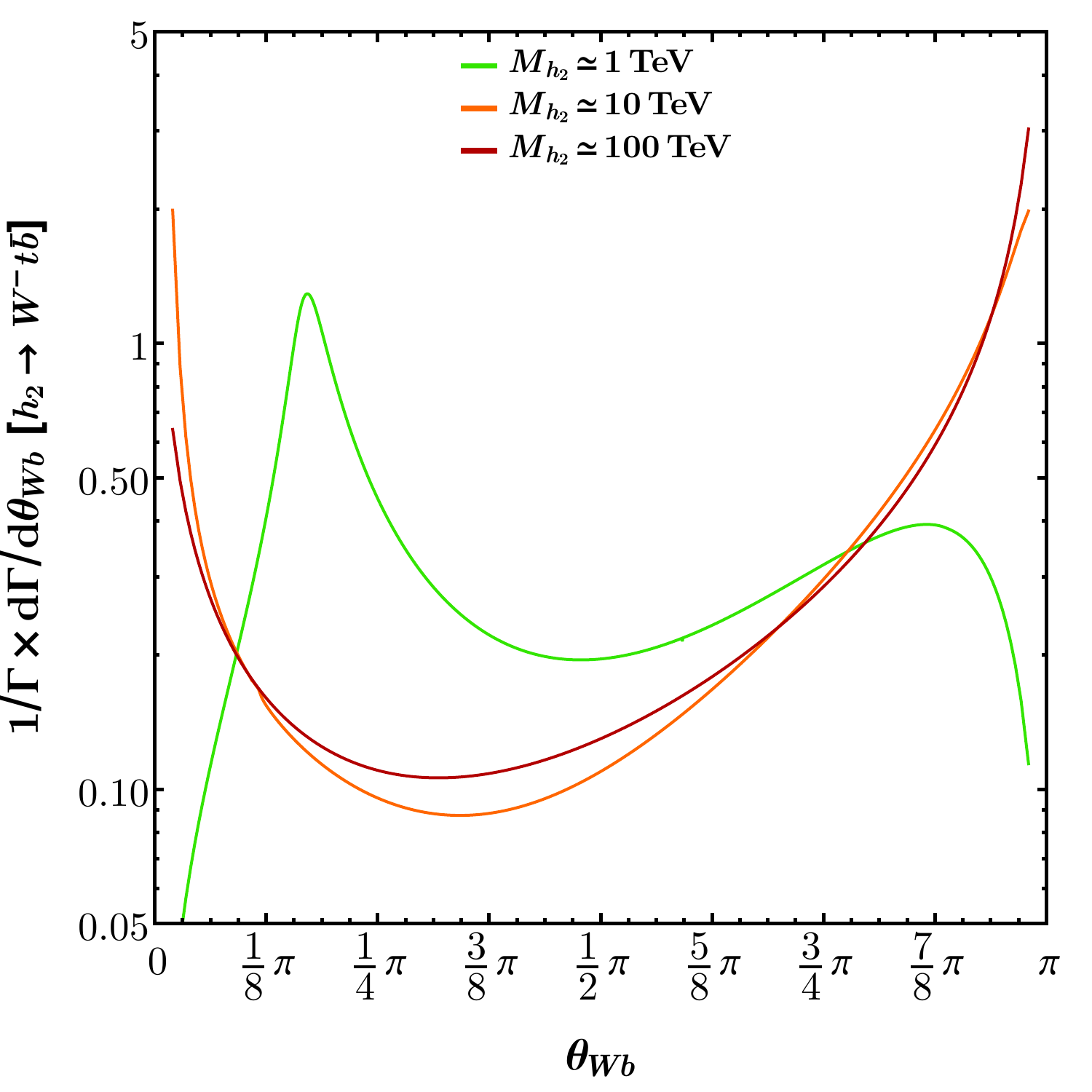}\\
  \includegraphics[width=0.48\textwidth]{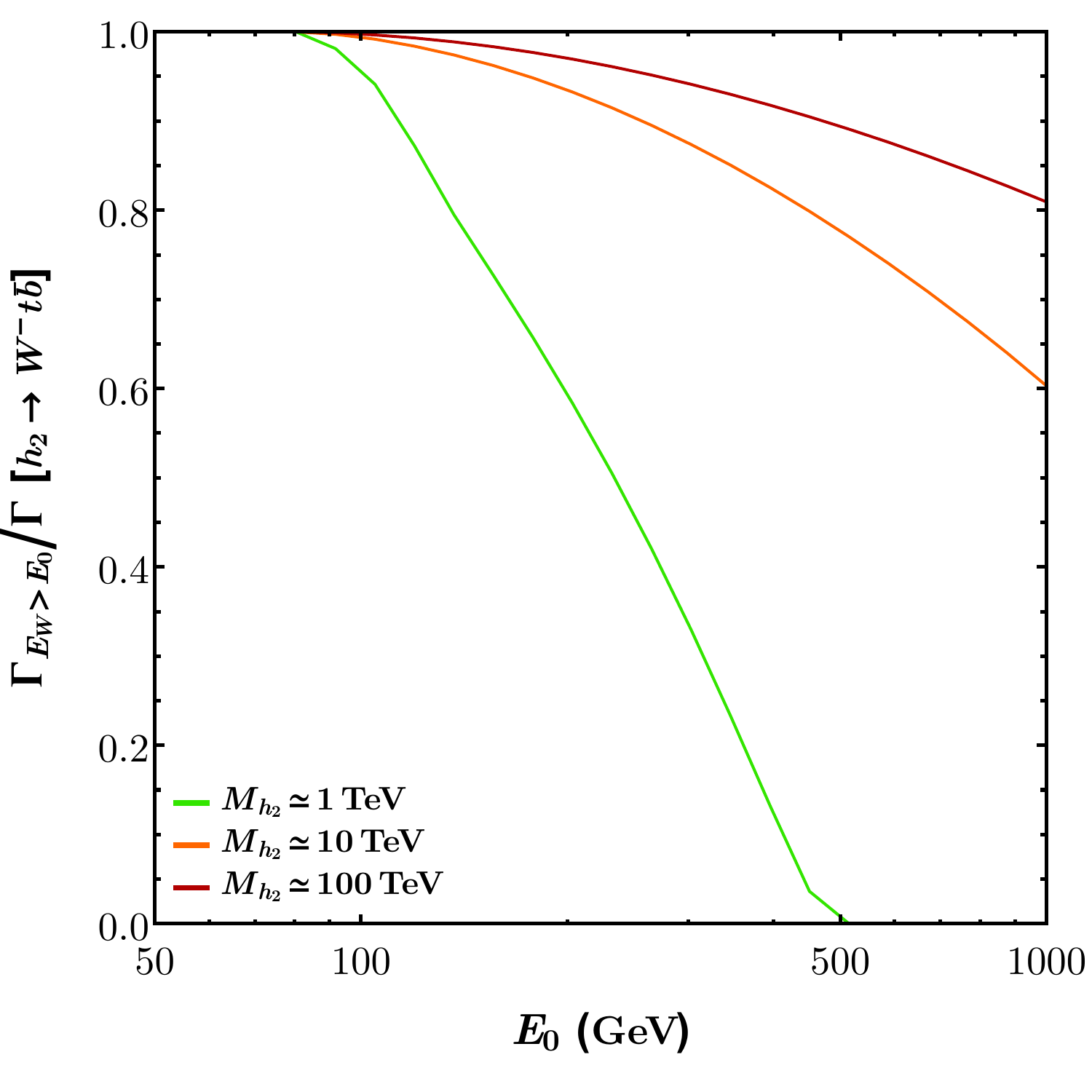}
  \includegraphics[width=0.48\textwidth]{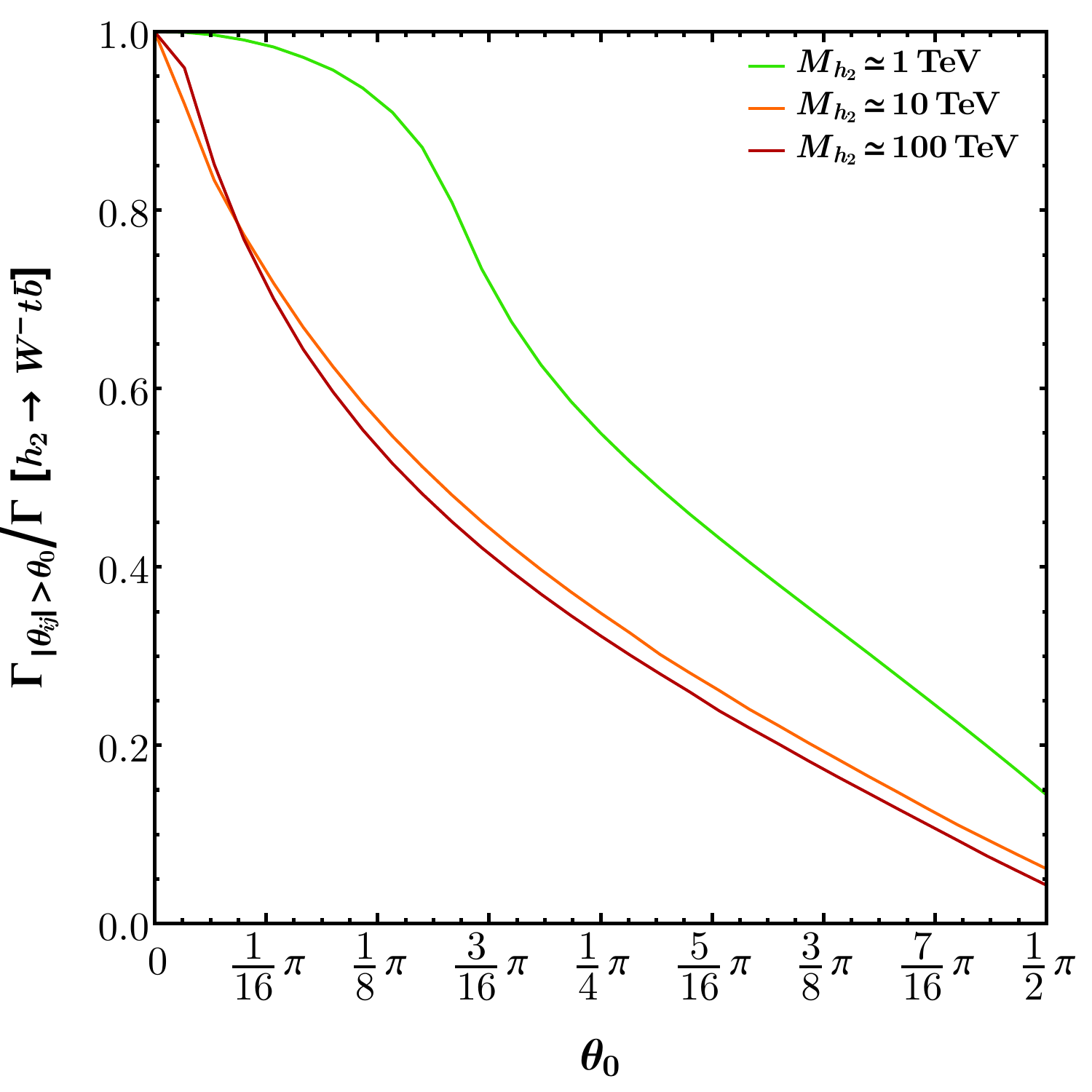}
  \caption{\label{fig:soft_col}
  Softness and collinearity of the off-shell contribution
  to~$\Gamma\left[\AtoB{h_2}{W^-t\bar{b}}\,\right]$
  for~\mbox{$M_{h_2}\approx1$}\,TeV~(green), $10$\,TeV~(orange)
  and~$100$\,TeV~(red).
  \newline\textit{Top left}: the differential cross-section in terms
  of the energy of the~$W$~boson in the rest-frame of~$h_2$
  systematically peaks in the lower range of available energy. The
  apparent spikes slightly above~\mbox{$E_W\sim0.1,1,10$}\,TeV
  (depending on the curve) are artifacts of the subtraction of the
  two-body
  contribution~\AtoB{h_2}{t\left(\AtoB{\bar{t}}{W\bar{b}}\right)}.
  \newline\textit{Bottom left}: the integrated cross-section as a
  function of the cutoff on the energy of the radiated $W$ boson.
  \newline\textit{Top right}: the differential cross-section in terms
  of the angle between the~$W$~boson and the~$b$~antiquark (in the
  rest-frame of~$h_2$) peaks for collinear emissions of~$W$
  and~$\bar{b}$ (at~\mbox{$\theta_{Wb}=0$}), or~$W$ and~$t$, or~$b$
  and~$\bar{t}$ (at~\mbox{$\theta_{Wb}=\pi$}).
  \newline\textit{Bottom right}: the fraction of the width captured
  after requiring a minimal angular separation of~$\theta_0$ between
  all the final states falls very rapidly at~\mbox{$\theta_0\gsim0$}.
  }
\end{figure}

From the IR~nature of the leading off-shell effects, one can expect
the (off-shell) three-body widths to exhibit soft and collinear
characteristics. We illustrate this fact in Fig.\,\ref{fig:soft_col}
for the channel~\AtoB{h_2}{W^-t\bar{b}}, with the three
cases~\mbox{$M_{h_2}\approx1$}\,TeV~(green), $10$\,TeV~(orange)
and~$100$\,TeV~(red). The differential cross-section is shown in the
upper row, in terms of the energy of the~$W$~boson~$E_W$~(left), and
in terms of the angle between the~$W$ and the~$b$
antiquark~$\theta_{Wb}$~(right), in the rest-frame of the Higgs
boson. We observe that the differential cross-section peaks at lower
values of~\mbox{$E_W\sim100\text{--}200$}\,GeV as well as
at~\mbox{$\theta_{Wb}\sim0$} (collinear~$W$ and~$\bar{b}$) or~$\pi$
(collinear~$W$ and~$t$, or~collinear~$t$ and~$\bar{b}$). The visible
spikes in the upper-left plot slightly
above~\mbox{$E_W\sim0.1,1,10$}\,TeV
for~\mbox{$M_{H^\pm}=1,10,100$}\,TeV are associated to the threshold
for the on-shell
decay~\AtoB{h_2}{t\left(\AtoB{\bar{t}}{W\bar{b}}\right)}; the
corresponding width is subtracted from the three-body width as a
Breit--Wigner distribution that, however, does not perfectly match the
threshold effect.

In the lower row of Fig.\,\ref{fig:soft_col}, we show how much of the
off-shell three-body decay width can be actually observed when
requiring a minimal energy of the radiated~$W$~(left), or when
requiring angular separations
of~\mbox{$\theta_0\in\left[0,\,\pi/2\right]$} between all the
particles in the final state (in the rest-frame of the decaying Higgs;
plot on the right): while the width depends only weakly on the energy
cutoff as long as the initial state is very heavy, up to~$40\%$ of the
signal could be lost with an angular cut
of~\mbox{$20^{\circ}\approx0.35$}
for~\mbox{$M_h\simeq10\text{--}100$}\,TeV. Thus, these three-body
decays may be experimentally challenging. In particular, the part of
the signal that produces collinear pairs with a~$W$~boson and
a~$b$~quark could be misinterpreted as originating in an
on-shell~$t$. On the other hand, a collinear~$t$--$W$~pair is unlikely
to be mistaken for an on-shell~$b$. The exact boundary between
two-body and three-body decays on the experimental side would probably
require a dedicated analysis, which goes beyond the scope of this
paper.

\begin{figure}[t!]
  \centering
  \includegraphics[width=0.46\textwidth]{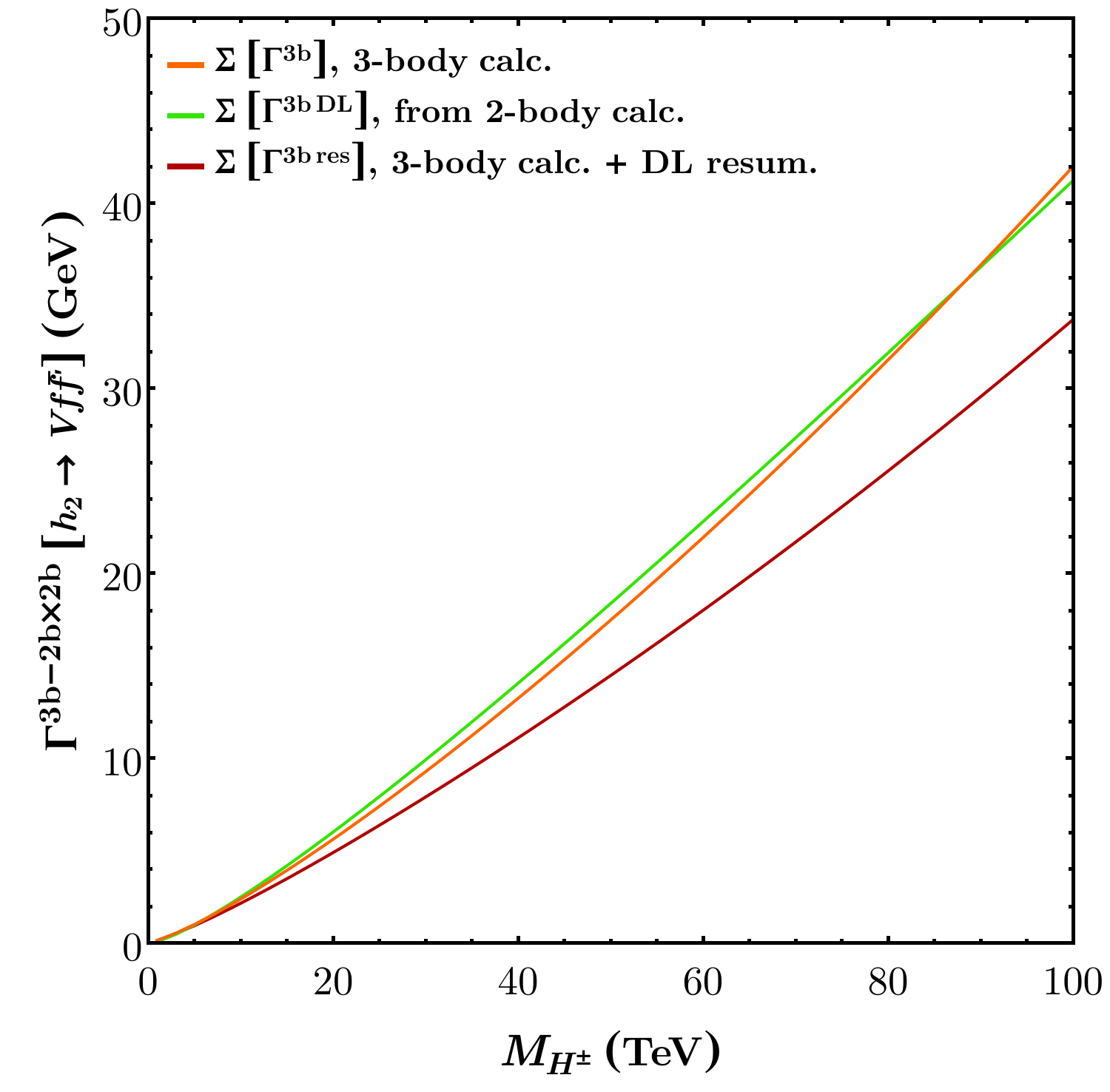}
  \includegraphics[width=0.46\textwidth]{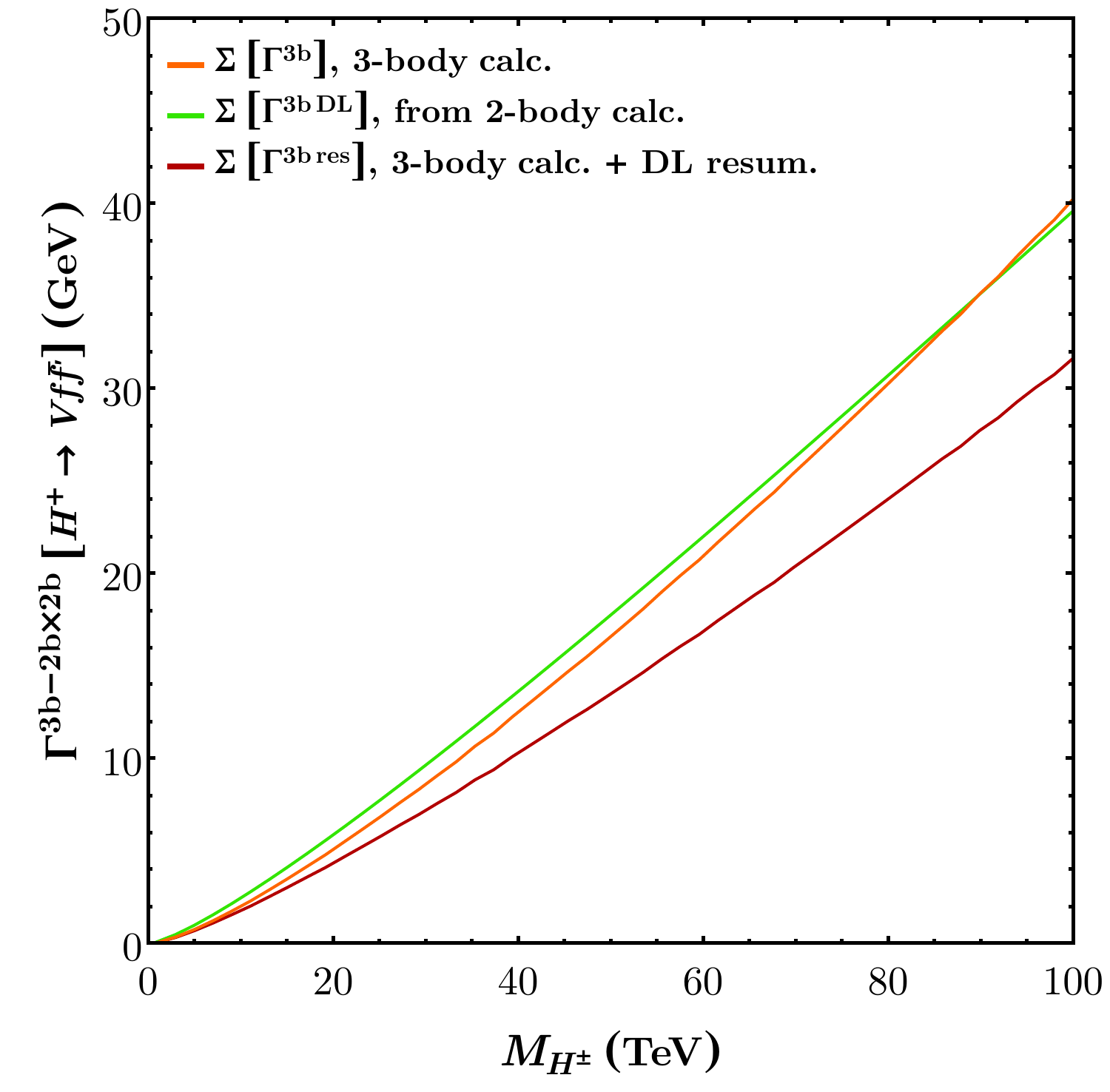}
  \caption{\label{fig:3b_resum}
  The inclusive three-body width and resummation of double~logarithms
  for a \cp-even doublet-like Higgs (left) and a charged Higgs (right)
  are shown. The sum of three-body contributions involving fermions of
  the third generation in the final state from a direct fixed-order
  calculation (as in Fig.\,\ref{fig:3b_width}, but also including the
  leptonic channels) is shown in orange. In green, we present the
  double-logarithmic approximation derived from the Sudakov
  double~logarithms of~\AtoB{h_2}{t\bar{t},b\bar{b},\tau^+\tau^-}; in
  red, we combine the fixed-order calculation and the resummation of
  double~logarithms.}
\end{figure}

The impact of the resummation of double~logarithms can be observed on
Fig.\,\ref{fig:3b_resum}. The green~curve corresponds to the
approximation of the inclusive three-body widths that can be inferred
from the Sudakov double~logarithms of the two-body
channels~\AtoB{h_2}{t\bar{t},b\bar{b},\tau^+\tau^-} (left)
and~\AtoB{H^+}{t\bar{b},\tau^+\nu_{\tau}} (right). It is roughly
comparable with the result obtained from a direct calculation of the
three-body channels~(orange~curve). The red~curve implements
Eq.\,\eqref{eq:res3bwidth}. That this curve is somewhat below the
green and orange~ones can be expected from its definition: if we
expand the exponential term in Eq.\,\eqref{eq:res3bwidth}, it is
obvious that we are adding a negative contribution from the
Sudakov~resummation to the widths obtained through explicit
calculation.

In these estimates of the three-body widths, we stress that (as our
educated guess) the same~QCD and QED~correction factors as in the
two-body widths are applied, in particular with QCD-running Yukawa
couplings at the scale of the mass of the decaying Higgs. While these
effects are formally of higher order, they have a sizable numerical
impact on the widths. At~\mbox{$M_{h_2}\approx100$}\,TeV for example,
the QCD~running represents a correction factor of about~$1/4$, meaning
that, evaluated with `pole' Yukawa couplings, the three-body widths
would dominate the two-body channels instead of being comparable. Yet,
the QCD$\times$EW order is in principle not under control in our
calculation, meaning that the large difference between the widths
obtained from `pole' and `running' Yukawa couplings should be
interpreted, strictly speaking, as a contribution to the theoretical
uncertainty.

\subsubsection*{Branching ratios}

The Higgs branching ratios are often more helpful quantities than the
decay widths, as they are more easily accessible experimentally. The
three-body decays then affect all decay channels through their
intervention in the full Higgs widths. We stress that the final states
including the radiation of electroweak gauge~bosons and Higgs bosons
are clearly distinguishable from the two-body final states and that
counting \EG~$W^-t\bar{b}$ as a~$b\bar{b}$~event is only weakly
motivated from an experimental perspective.\footnote{On the other
hand, part of the signal~\AtoB{h_2}{W^-t\bar{b}\,} could be counted
as~\AtoB{h_2}{t\bar{t}\,} events. However, for the considered value
of~$\tan\beta$, most of the contributions from radiated gauge~bosons
are actually associated with the Yukawa coupling of the bottom and
would thus be more appropriately attached
to~$\Gamma\left[\AtoB{h_2}{b\bar{b}\,}\right]$ in an inclusive
analysis.} We thus define the branching fractions as actual ratios of
the two-body decay widths by the total width.

\begin{figure}[tp!]
  \centering
  \includegraphics[width=0.48\textwidth]{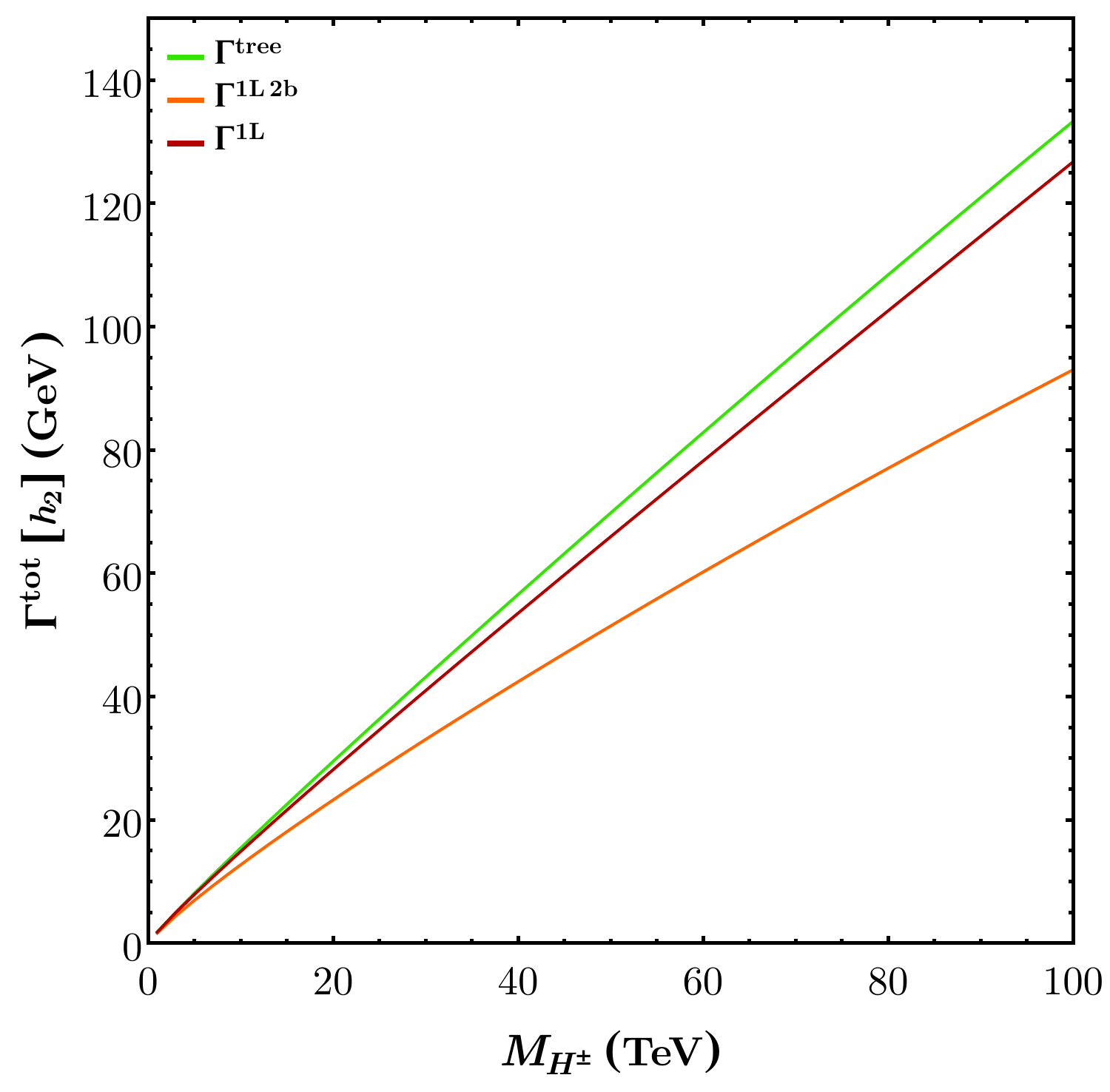}
  \includegraphics[width=0.48\textwidth]{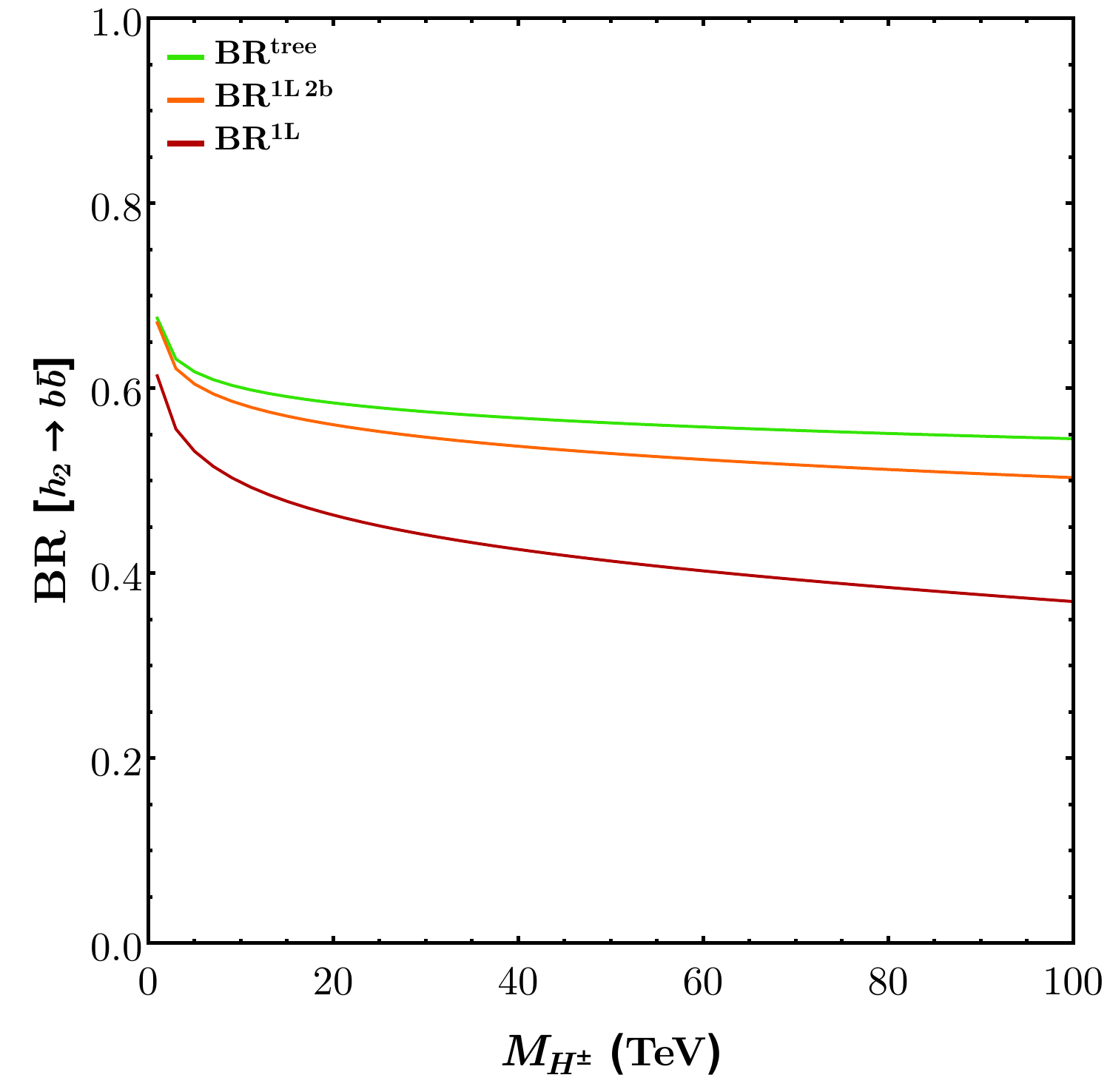}\\
  \includegraphics[width=0.48\textwidth]{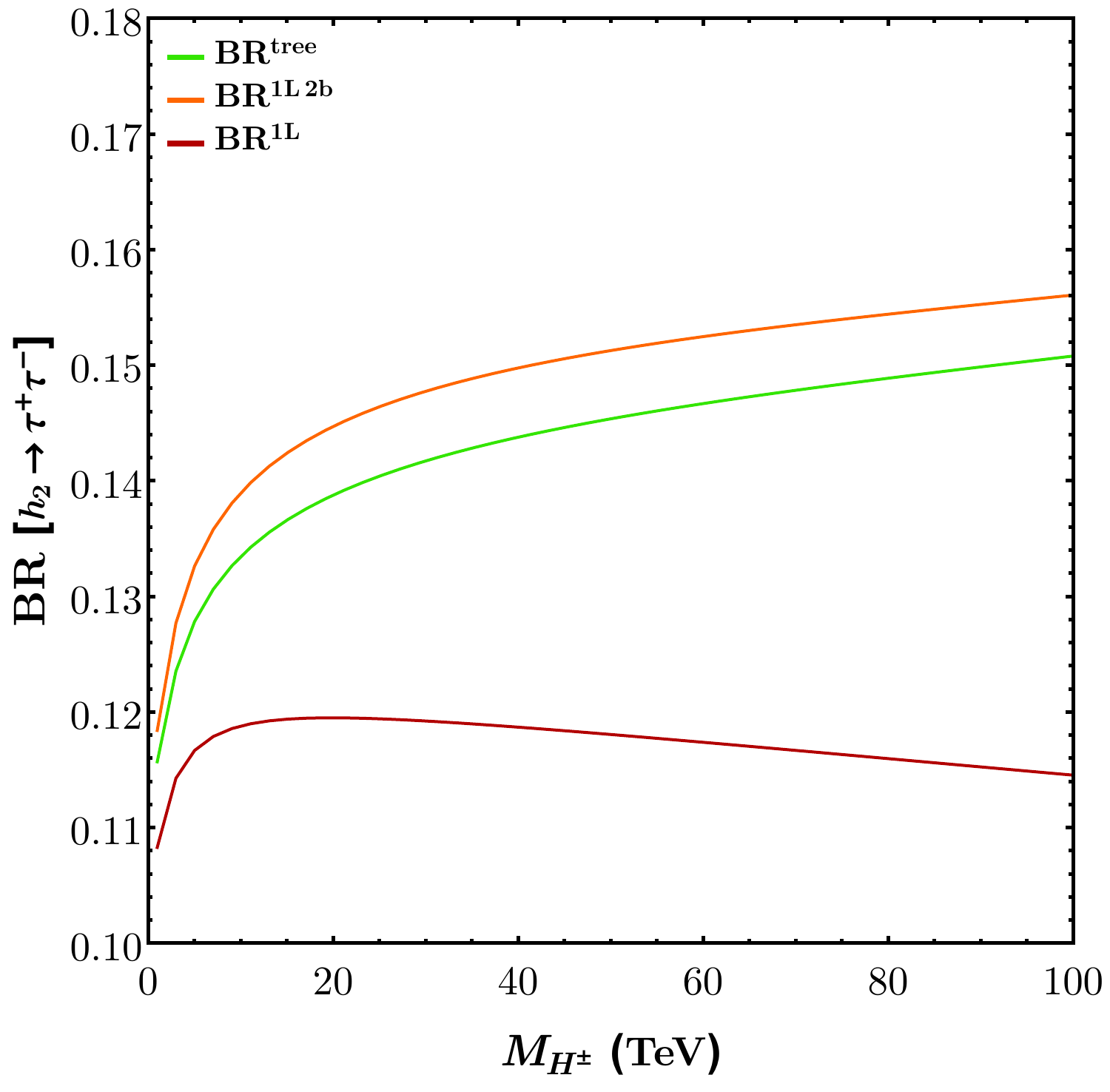}
  \includegraphics[width=0.48\textwidth]{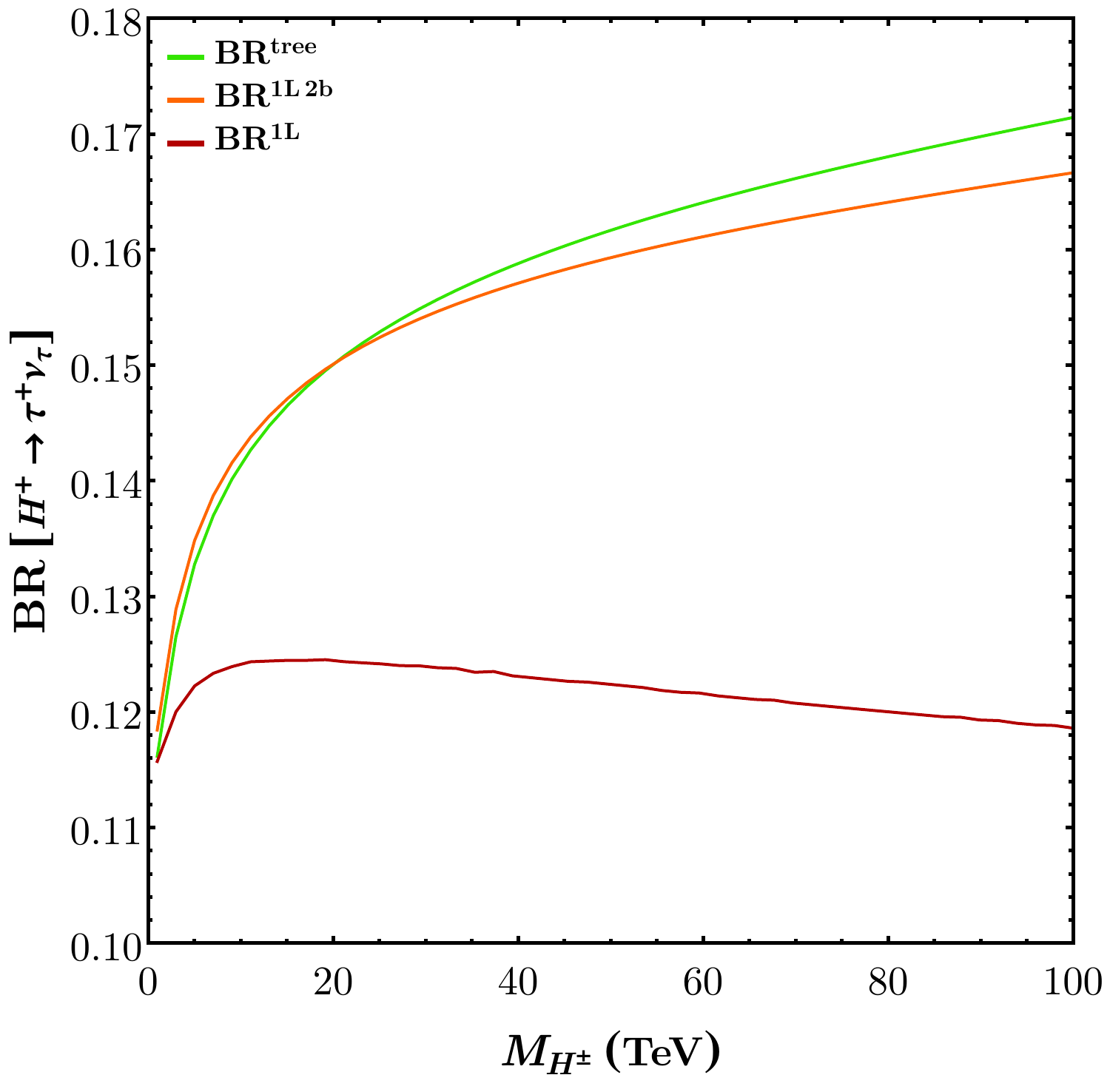}
  \caption{\label{fig:BRs}
  Total width and branching ratios of a heavy doublet-like Higgs
  states: \newline\textit{Top left}: total width of the \cp-even
  state~$h_2$ at the (QCD/QED/SUSY-corrected) tree-level
  order~(green), considering only the two-body decays at one-loop
  order~(orange), and including also the three-body
  decays~(red).  \newline\textit{Top right}: branching ratio of
  the \cp-even state~$h_2$ into~$b\bar{b}$.  \newline\textit{Bottom
  left}: branching ratio of the~\cp-even state~$h_2$
  into~$\tau^+\tau^-$.  \newline\textit{Bottom right}: branching ratio
  of the charged state~$H^+$ into~$\tau^+\nu_{\tau}$.  \newline The
  branching ratios are derived at the same orders as the total
  widths, \IE~employing only (QCD/QED/SUSY-corrected) effective
  widths~(green), considering only the two-body decays at the one-loop
  order~(orange), and finally, including also the three-body
  decays~(red).}
\end{figure}

We continue to consider the decoupling scenario that has been
presented before and show the results in Fig.\,\ref{fig:BRs}. The
following three approximations are displayed: in green, only the
QCD/QED-corrected widths~$\Gamma^{\text{eff}}$ (together with the
factorized SUSY~effect) are included, thus dismissing the electroweak
corrections; the orange~curves are obtained by considering the
two-body widths~$\Gamma^{\text{imp}}$ at the full one-loop order, but
no three-body channel is counted in the total width (thus amounting to
an incomplete one-loop result); in red, we add the three-body decay
widths in the total width, thus achieving a consistent prediction of
the two-body branching fractions at the full one-loop order. In all
these evaluations, the fermionic decays are implemented as described
in Sect.\,\ref{sec:theory}, including in particular the resummation of
Sudakov double~logarithms. In addition, the two-body decays into the
gauge~bosons of the~SM are incorporated at the (improved) one-loop
order, as defined in \citere{Domingo:2018uim}. The decays involving
final states with electroweak gauge~bosons are suppressed at the
tree~level in the decoupling limit, so that corrections of
Sudakov~type play a negligible part in radiative effects (dominated by
the fermion loops). All other two-body decays are included at the tree
level, in particular the suppressed~\AtoB{h_2}{Vh_1}
or~\AtoB{h_2}{2\,h_1}~channels (decays into SUSY~particles are
kinematically forbidden in the scenario under consideration).

On the left-hand side of Fig.\,\ref{fig:BRs}, in the upper row, we
first display the total widths of the heavy~\cp-even state for the
three considered approximations: the one-loop corrections to the
two-body decays (orange~curve) lead to a suppression of the total
width (as compared to the effective tree-level order, in green),
consistent with what was observed in Fig.\,\ref{fig:hbb_width}, and
mainly driven by the Sudakov double~logarithms of
Eq.\,\eqref{eq:2bDL}. Nevertheless, the inclusion of the three-body
decays~(red) approximately restores the magnitude of the
(QCD/QED-corrected) tree-level total width: this is an expected
result, due to the cancellation of the double~logarithms between the
two- and three-body decays. On the right-hand side, we show the
branching
ratio~$\text{BR}{\left[\AtoB{h_2}{b\bar{b}}\,\right]}$, \IE~that of
the naively dominant decay channel of the \cp-even state. We observe
that this quantity is suppressed, as compared to the tree-level
prediction, by the decrease of the two-body width~(orange line) but
even more so by the inclusion of the three-body decays in the total
width~(red line). The impact of the one-loop corrections to the
two-body decays is only at the percent level, while that of the
inclusion of the three-body decays in the total width is of
order~$\simord10\%$ at~\mbox{$M_{h_2}\approx1$}\,TeV and
reaches~$\simord30\%$ at~\mbox{$M_{h_2}\approx100$}\,TeV. This can be
easily understood when considering that the total width in the strict
two-body approximation is suppressed in about the same proportion by
the double~logarithms as~$\Gamma\left[\AtoB{h_2}{b\bar{b}}\,\right]$
(its leading contribution): therefore, we expect little difference
between the green and orange curves. On the contrary, when we include
the three-body decays, the total width is approximately restored to
its effective tree-level value, so
that~$\text{BR}{\left[\AtoB{h_2}{b\bar{b}}\,\right]}$ becomes
sensitive to the suppression of the associated decay width through
double~logarithms.

In the second row of Fig.\,\ref{fig:BRs}, we turn to the leptonic
decays~\AtoB{h_2}{\tau^+\tau^-}~(left)
and~\AtoB{H^+}{\tau^+\nu_{\tau}} (right) as the typical search
channels at proton--proton colliders. Contrarily to what we observed
in the case of~$\text{BR}{\left[\AtoB{h_2}{b\bar{b}}\,\right]}$, the
prediction for~$\text{BR}{\left[\AtoB{h_2}{\tau^+\tau^-}\right]}$ that
is exclusively based on one-loop-corrected two-body widths~(orange) is
not suppressed, but enhanced. Indeed, in this approximation, the total
width, dominated by~\AtoB{h_2}{b\bar{b}}, receives a larger
suppression as compared to the width of~\AtoB{h_2}{\tau^+\tau^-}: the
effect amounts to less than~$\simord5\%$ depending on~$M_{h_2}$, and
the green and orange~curves show the same behavior---and so do they in
the case~$\AtoB{H^+}{\tau^+\nu_{\tau}}$. However, the estimate at the
full one-loop order (\IE\ including the three-body decays; in red)
shows the opposite trend. Again, as the three-body channels restore
the magnitude of the total width to approximately its original
effective value, there cannot be an associated enhancement any
longer. On the contrary, the branching fractions become sensitive to
the suppression of the tau-onic widths via Sudakov
effects. Consequently, the discrepancy with the two-body approximation
is sizable, loosely amounting to about~$10\%$ of the corresponding
branching ratio at~\mbox{$M_{h_2}\approx1$}\,TeV and
reaching~$\simord30\%$ at~\mbox{$M_{h_2}\approx100$}\,TeV.

This short study shows that it is in fact misleading to assume that
the one-loop corrections at the level of the two-body widths are
sufficient to provide predictive branching ratios at the full one-loop
order. We actually observe little difference with the effective
tree-level in this approximation, while the inclusion of the
three-body decays generates a sizable shift.

%% file: 04_Conclusions.tex
\tocsection[\label{sec:conclusion}]{Conclusions}

In this paper, we have emphasized the impact of the electroweak
radiative corrections on the decays of heavy Higgs states. Focusing on
the specific case of fermionic decays in the NMSSM, we have shown that
such contributions are dominated by effects of IR~type, namely Sudakov
double~logarithms, which could have a sizable impact on the magnitude
of the two-body decay widths. This structure also provides a grasp on
the electroweak corrections of higher order, since the leading
double~logarithms can be explicitly resummed.

In addition, we have stressed the relevance of three-body decay
channels for a consistent evaluation of the total widths and branching
ratios at the full one-loop order. The IR~effects of the two-body
widths are indeed mirrored by double~logarithms in the radiation of
electroweak gauge~bosons, whereas the emission of light Higgs bosons
only contributes at single-logarithmic order. Contrarily to the
radiation of gluons and photons in~QCD/QED, the radiation of
electroweak gauge~bosons leads to clearly distinguishable final states
(due to the explicit IR~cutoff fulfilled by the mass of the
gauge~bosons) and, in any case, the definition of an inclusive width
is non-trivial (\EG\ due to the flavor changes associated to emissions
of~$W$~bosons). On the other hand, the off-shell three-body signals
exhibit the expected soft and collinear characteristics that could
make them difficult to extract experimentally.

Numerically, the inclusion of the three-body decays largely
compensates the suppression of the two-body widths associated to
electroweak effects, resulting in an approximately stable total width
when simultaneously including or discarding the electroweak
virtual corrections and real emissions. As a
consequence, the branching ratios into two-body final states are fully
sensitive to the suppression of the two-body widths associated to the
double-logarithmic effects. We insist upon our conclusion that an
estimation of the total Higgs widths and the branching ratios at the
one-loop order solely from the loop-corrected two-body widths fails to
capture the actual impact of the electroweak corrections.

In this work, we have focused on final states with SM~fermions, since
they are the expectedly leading decay channels of heavy doublet-like
Higgs bosons involving SM particles in the final state---bosonic
decays are suppressed in the decoupling limit. However, the main
ingredients that we have discussed here also apply in alternative
scenarios, \EG~decays into bosonic or SUSY final states. Indeed,
three-body channels always need to be considered in parallel with
one-loop corrections for a consistent order-counting and the
Sudakov~effects are always expected to emerge in electroweak
corrections as long as there exists a hierarchy between the mass of
the decaying Higgs (center-of-mass energy) and the decay
products. However, these effects are only significant if the
born-level amplitude itself is unsuppressed. By lowering the scale of
the SUSY~spectrum such that heavy Higgs bosons can decay into
(comparatively) light pairs of neutralinos, charginos or sfermions, we
should thus again recover sizable Sudakov~corrections and relevant
three-body decay widths in these channels with SUSY~final states.